\documentclass{aa}  
\usepackage{graphicx}
\usepackage{txfonts}
\usepackage{ulem}
\usepackage{graphicx}
\usepackage{tabularx}
\usepackage{txfonts}
\usepackage{longtable}
\usepackage{natbib}
\usepackage{hyperref}
\usepackage{verbatim}
\usepackage{url}
\usepackage{float,color}
\usepackage{subcaption}
\usepackage[figuresright]{rotating}
\hypersetup{
    colorlinks=true,
    linkcolor=blue,
    filecolor=magenta,      
    urlcolor=cyan,
    citecolor=blue
}

\bibpunct{(}{)}{;}{a}{}{,}
\newcommand{\uproman}[1]{\uppercase\expandafter{\romannumeral#1}}

\newcommand{\logg}{$\rm \log{g}$}

\newcommand{\dis}{\texttt{DISPATCH}\ }
\newcommand{\medis}{\texttt{M3DIS}}

\newif\ifhighlight
\highlightfalse 

\newcommand{\revisedPE}[1]{\ifhighlight\textbf{\textcolor{red}{#1}}\else#1\fi}

\begin{document}

   \title{M3DIS - A grid of 3D radiation-hydrodynamics stellar atmosphere models for stellar surveys}
\titlerunning{M3DIS: A grid of 3D RHD stellar atmosphere models for stellar surveys}

   \subtitle{II. Carbon-enhanced metal-poor stars{\thanks{All \medis\ 3D RHD CEMP models are available under \url{https://nlte.mpia.de}}}}

   \author{Philipp Eitner \inst{1,2} \thanks{E-mail: eitner@mpia.de} \and  Maria Bergemann \inst{1,3} \and Richard Hoppe \inst{1,2} \and Nicholas Storm \inst{1,2} \and Veronika Lipatova \inst{4,2} \and\\ Simon C. O. Glover \inst{4} \and Ralf S. Klessen \inst{4,5,8,9} \and {\AA}ke Nordlund \inst{3,6} \and Andrius Popovas\inst{6,7}}

    \institute{
    Max-Planck-Institut f{\"u}r Astronomie, 69117 Heidelberg, Germany
    \label{1}
    \and
    Universit{\"a}t Heidelberg, Grabengasse 1, 69117 Heidelberg, Germany
    \label{2}
    \and
    Niels Bohr International Academy, Niels Bohr Institute, University of Copenhagen, DK-2100 Copenhagen, Denmark
    \label{3}
    \and 
    Universit{\"a}t Heidelberg, Zentrum für Astronomie, Institut für Theoretische Astrophysik, Heidelberg, Germany
    \label{4}
    \and 
    Universit{\"a}t Heidelberg, Interdisziplin{\"a}res Zentrum für Wissenschaftliches Rechnen, Heidelberg, Germany
    \label{5}
    \and
    Rosseland Centre for Solar Physics, University of Oslo, P.O. Box 1029, Blindern, NO-0315 Oslo, Norway
    \label{6}
    \and
    Institute of Theoretical Astrophysics, University of Oslo, P.O. Box 1029, Blindern, NO-0315 Oslo, Norway
    \label{7}
    \and
    Harvard-Smithsonian Center for Astrophysics, Cambridge, USA
    \label{8}
    \and 
    Elizabeth S. and Richard M. Cashin Fellow at the Radcliffe Institute for Advanced Studies at Harvard University, Cambridge, USA
    \label{9}
    }
    \date{}

  \abstract
   {}
   {Understanding the origin and evolution of carbon-enhanced metal-poor (CEMP) stars is key to tracing the early chemical enrichment of the Galaxy. In this work we investigate how physically realistic 3D radiation-hydrodynamic (RHD) carbon-enhanced model atmospheres affect the inferred carbon abundances in CEMP stars, and assess the implications for their classification and for Galactic chemical evolution (GCE). We pay particular attention to the systematic biases introduced by traditional 1D hydrostatic equilibrium (HE) models.}
   {We used the \medis\ code to compute 3D RHD model atmospheres for main-sequence and sub-giant stars spanning a wide range of metallicities and carbon enhancements. Synthetic spectra of the CH G band were calculated using full 3D radiative transfer and compared to spectra from classical 1D HE MARCS models. We derived abundance corrections and applied them to a large literature sample of metal-poor stars from the SAGA database to quantify systematic effects on the carbon abundance distribution and CEMP classification.}
   {Our new 3D CEMP models predict significantly cooler upper atmospheric layers than 1D HE models, resulting in stronger CH absorption and lower inferred carbon abundances by up to $-0.9$ dex at the lowest metallicities. Carbon enhancement in the atmosphere itself increases molecular opacities and leads to radiative re-heating, which partly offsets the adiabatic cooling in 3D models and reduces the magnitude of $\rm 3D-1D$ abundance corrections. Applying these corrections lowers the CEMP fraction by up to $20\%$ below $\rm [Fe/H] = -3$, and furthermore alters the relative contribution of CEMP sub-classes. In particular, the fraction of stars classified as CEMP-no  increases and that of CEMP-r/s stars decreases, owing to the downward revision of absolute carbon abundances. These changes bring the Galactic carbon abundance distribution into better agreement with GCE models that assume a contribution from faint supernovae of $20\%$. Physically realistic model atmospheres are thus essential for a reliable reconstruction of the early chemical enrichment history of the Galaxy.}
   {}


\maketitle
%
\section{Introduction}\label{sec:introduction}

Carbon-enhanced metal-poor (CEMP) stars are of major interest in modern astrophysics. Chemical abundances measured in the spectra of these stars are relevant for understanding the properties of type II supernova (SN)  events in the early Galaxy, nucleosynthesis,  the formation and evolution of binary stars, the stellar initial mass function, the metallicity distribution function of the Galactic halo, and the production sites of neutron-capture elements \citep{Beers2005,Aoki2006,Hansen2018}.

Carbon-enhanced metal-poor  stars are defined based on the amount of carbon (C) compared to iron (Fe) relative to the solar composition present in their atmospheres. Originally, the limit for CEMP classification was placed at $\rm [C/Fe]>1.0$ \citep{Beers2005}, which was later slightly adjusted to the presently applied criterion of $\rm[C/Fe]>0.7$ \citep{Aoki2007}.
It is furthermore common to subclassify CEMP stars based on the presence of elements synthesised in the slow or rapid neutron capture processes \citep[s- and r-processes;][]{Hirschi2006}. The categories include stars with enhancements in one of the two channels (CEMP-s and CEMP-r), in both (CEMP-r/s), or in neither (CEMP-no), and each category is thought to probe a different formation environment \citep[e.g.][]{Masseron2010}. CEMP-s stars are viewed as products of enrichment by a post-asymptotic giant branch companion in a binary system \citep{McClure1990,Lucatello2005,Starkenburg2013,Hansen2016b}, although some CEMP-s stars appear to be single, which might indicate that they formed from the ejecta of a massive, fast rotating source star with a late mixing event \citep{Choplin2017}. In contrast, CEMP-r stars comprise a larger \revisedPE{enhancement} of r-process elements, such as Eu \citep{Sneden2003}, and are thought to form from matter that became chemically polluted by a core-collapse supernova (CCSN; electron-capture or magneto-rotating), a neutron star--neutron star merger, or a neutron star--black hole merger \citep{Frebel2015,cote2017,Hansen2018,Kobayashi2020}.

The formation history of CEMP-r/s and CEMP-no stars is less clear. Regarding CEMP-r/s, there is evidence of an r-process contribution \citep{Aoki2006} from a SN II. Alternative explanations include s-process enrichment  by a thermally pulsating asymptotic giant branch companion \citep{Masseron2006} and neutrino-driven winds from the accretion-induced collapse of an O-Ne-Mg white dwarf  \citep{Cohen2003,Jonsell2006,Kobayashi2020}. CEMP-no stars are among the most metal-poor stars observed \citep{Christlieb2002,HansenT2016} and show no significant enrichment in either the r- or s-process \citep{Spite2013}. This can be interpreted in the framework of different scenarios, including nucleosynthesis in faint low-energy SNe \citep{Umeda2005,Tominaga2007,Nomoto2013,Kobayashi2020}, hypernovae \citep{Heger2002,Limongi2003,Nomoto2013}, nucleosynthesis in C-rich winds of rotating massive stars \citep{Hirschi2006,Meynet2006}, and early-asymptotic giant branch stars \citep[][and references therein]{Masseron2010}.

The analysis of C abundances in the atmospheres of CEMP stars primarily rests upon spectroscopic diagnostic of the so-called G band \citep[e.g.][]{Placco2016a,Placco2016b,francois2018,Norris2019}. The G band represents the fundamental A-X transition of the CH molecule \citep{Masseron2014,Popa2023} and is therefore highly sensitive to the abundance of C and to the temperature gradient in stellar atmospheres. The G band has been extensively studied using 1D local thermodynamic equilibrium (LTE) models \cite[e.g.][see also \citealp{Suda2008,Suda2017} and references therein]{Lai2007,Roederer2014,Jacobson2015,Lee2013}, but recent analyses have also explored the formation of the G band in 3D radiation-hydrodynamics (RHD) model atmospheres of metal-poor dwarfs and red giants 
\citep[e.g.][]{Collet2007,Bonifacio2013,Gallagher2016, Gallagher2017,Collet2018}.

In this  study we investigated the effect of carbon enhancement on the structure of stellar atmospheres in realistic 3D RHD simulations introduced in our previous work \citep{Eitner2024}. We further studied and quantified the impact of 3D convection on the diagnostics of C abundances from stellar spectra. We then analysed the results with respect to CEMP statistics in the Galaxy and discuss implications for the formation and evolution of CEMP stars.

The paper is organised as follows. In Sect. \ref{sec:methods} we briefly introduce the stellar observations used in this work, followed by a description of the \medis\ code and model parameters. In Sect. \ref{sec:results} we present our new 3D RHD model atmosphere calculations, including a comparison with 1D hydrostatic equilibrium (HE) MARCS models, the spectrum synthesis of the CH G band between $4297\,$\AA\ and $4303\,$\AA, and the analysis of CEMP abundances in 3D, along with a discussion of their implications for Galactic CEMP populations. Finally, in Sect. \ref{sec:conclusions} we summarise our findings and discuss their implications for the study of CEMP stars and Galactic chemical evolution (GCE).

\section{Methods} \label{sec:methods}
\begin{table}
\renewcommand{\arraystretch}{1.2}  
\caption[]{\texttt{M3DIS} RHD code features.}
\label{tab:code}
\centering
\begin{tabular}{p{3cm} p{4.5cm}}
\hline\hline                 
Feature & Description \\  
\hline
Code Name & \texttt{M3DIS} -- \textbf{M}odels in \textbf{3}D @\textbf{DIS}PATCH \citep{Eitner2024,Nordlund2018} \\
Dimensionality & Full 3D \\
Physics Included & Radiation and convection \\
Applications & FGKM stars; any $\log g$; any metallicity \\
Opacity Sources & 92 elements; electron and Rayleigh scattering; molecules \\
Boundary Conditions & Top: Open, in-/outflows, exponential $\rm \rho$ drop, constant $\rm e/m$; Bottom: Open, outflows free, inflows maintain entropy (via ghost zones) \\
Time Integration & Explicit; local time step based on the Courant–Friedrichs–Lewy (CFL) condition from gas velocity, radiative heating and radiative diffusion \\
Grid Type & Eulerian; uniform in all directions \\
Hydrodynamics Solver & HLLC Riemann solver \citep{Fromang2006} \\
Radiative Transfer & Short-characteristics; conservative on staggered mesh; 8 frequency bins; 4 $\rm \phi$, 2 $\rm \mu$ angles, plus disk-centre rays \\
\hline
\end{tabular}
\end{table}
\subsection{Stellar observations}  \label{subsec:data}
In this study we relied on stellar C abundances provided in the Stellar Abundances for Galactic Archaeology (SAGA) database\footnote{\url{http://sagadatabase.jp/}} \citep[][]{Suda2008,Suda2011,Suda2017}. SAGA is a comprehensive source of abundances and stellar parameters for stars in the Milky Way and its satellites, with a particular focus on metal-poor stars. The data are based on published, peer-reviewed sources and it has a particular focus on homogeneity of measurements, such that abundances are derived from 1D LTE synthetic spectra \citep{Suda2008}. 

The focus of this study was main-sequence (MS) and sub-giant stars. Hence, we did not include stars with a surface gravity $\rm log(g) < 2.7$. For each star, we extracted stellar parameters (the effective temperature and \logg), as well as [Fe/H] and [C/Fe]. Since Eu and Ba abundances are available for a very limited number of stars, further classification into CEMP-no, s and r/s is not straightforward. However, as shown by \cite{Yoon2018}, classification into group \uproman{1}, \uproman{2} and \uproman{3} is still possible based on [C/Fe] and [Fe/H] alone, which then allows an approximate separation of CEMP-no stars from those that are enhanced in n-capture elements.

In total, our sample contains $3192$ stars with sub-solar metallicities, [Fe/H] $< 0$, out of which $1745$ have a measured [C/Fe] abundance, with $541$ exceeding $\rm [C/Fe]\geq0.7$ and hence can be classified as CEMP \citep{Yoon2018}. The references to individual datasets along with stellar parameters and abundances (original and $\rm 3D-corrected$ values) are provided as supplementary material via the CDS. 

\subsection{Model atmospheres}  \label{subsec:model_atmospheres}
In this work we relied on the \texttt{M3DIS} (Models in 3D@\texttt{DISPATCH}) code within the \dis framework \citep{Nordlund2018} for generating 3D RHD stellar atmospheres, which has been introduced and extensively tested by \cite{Eitner2024}. In short: \medis\ solves the equations of hydrodynamics together with the radiative transfer (RT) equation to compute the radiative cooling, which in turn contributes to the energy balance of the atmosphere. The code relies on a HLLC Riemann solver for advancing the simulation in time \citep{Fromang2006}, and on short-characteristics RT in two $\rm \mu$ (polar) and four $\rm \phi$ (azimuthal) angles, plus upward and downward vertical rays, assembled in a Gauss-Radau quadrature. The main features of the models can be found in Table \ref{tab:code}.
\begin{table}
\setlength{\tabcolsep}{4pt}
\caption[]{Stellar parameters and carbon abundances of our 3D CEMP model atmospheres.}  
\label{tab:models}  
\centering
\begin{tabular}{ccccccc}  
\hline\hline                 
$\rm T_{eff}$ & $\rm log (g)$ & $\rm [Fe/H]$ & $\rm [C/Fe]$ & $\rm A(C)$ & Resolution & Size \\  
$\rm K$ & $\rm dex$ & $\rm dex$ & $\rm dex$ & $\rm dex$ & $\rm N_x\times N_y\times N_z$ & Mm \\ 
\hline                        
  5750 & 4.5 & -6 & 4 & 6.56 & $240^2\times120$ & 5.2 \\
  5750 & 4.5 & -5 & 3 & 6.56 & $240^2\times120$  & 5.2 \\
  5750 & 4.5 & -4 & 2 & 6.56 & $240^2\times120$  & 5.1 \\
  5750 & 4.5 & -3 & 1 & 6.56 & $240^2\times120$  & 5.2\\
  5750 & 4.5 & -2 & 1 & 7.56 & $240^2\times120$  & 5.4\\
  5250 & 3.0 & -6 & 4 & 6.56 & $240^2\times120$  & 140 \\
  5250 & 3.0 & -5 & 3 & 6.56 & $240^2\times120$  & 140 \\
  5250 & 3.0 & -4 & 2 & 6.56 & $240^2\times120$  & 140 \\
  5250 & 3.0 & -3 & 1 & 6.56 & $240^2\times120$  & 150 \\
  5250 & 3.0 & -2 & 1 & 7.56 & $240^2\times120$  & 160 \\
\hline                                   
\end{tabular}
\tablefoot{
The level of C-enhancement varies from $\rm [C/Fe] = 1$ (at [Fe/H] $=-3$) to $\rm [C/Fe] =4$ (at [Fe/H] $=-6$). Here, $\rm N_i$ corresponds to the number of grid points in dimension $\rm i$. In the last column, we provide the horizontal extend of the corresponding model atmosphere. Note that the vertical extend is half that number. For every CEMP model listed here, we generated a scaled-solar ($\rm[C/Fe]=0$) counterpart. For each model, convergence was achieved in less than $3500$ CPU-h.
}
\end{table}

The RT is solved in the multi-group approximation \citep{Nordlund1982,Ludwig1990}, such that the monochromatic opacities and source functions are collected into eight bins based on wavelength and formation height within the respective atmosphere. The selection of opacity bin edges is done automatically using a generic k-means clustering approach. Monochromatic opacities are computed together with the internal energy and gas pressure, i.e. the equation of state (EoS), on a grid of $\rm 151 \times 151$ $\rm \log \rho - \log T$ points using the updated version of \texttt{MULTI3D} \citep{Leenaarts2009}, also within the \dis framework. Scattering due to free electrons and Rayleigh scattering (H I, H$_2$, and He I) is treated as true absorption in the optically thick-part of the opacity table, and it is not excluded in the optically thin regime. This is a common approach in 1D \citep{Gerber2023} and 3D \citep{Collet2011} calculations. Opacities and source functions are computed under the assumption of LTE in the context of computing surface cooling. For more details, see the description by \cite{Eitner2024} and Appendix \ref{appendix:supp_material}.

In addition to its computation speed, major advantages of \medis\ are the highly flexible and consistent opacities and EoS, which make the simulation of stars with non-solar chemical composition relatively straightforward. To investigate the impact of the carbon abundance on the atmospheric structure itself, we first computed the scaled-solar $\rm [C/Fe]$ composition \citep[relative to][]{Magg2022} opacity and EoS tables at $\rm [Fe/H] = -6,-5,-4,-3,\ and\ -2$. Second, we calculated carbon-enhanced opacity tables, such that at $\rm[Fe/H] \leq -3$ we additionally have C-enhanced atmospheres with the same, constant $\rm A(C) = 6.56$, while at $\rm [Fe/H] =
-2$ we keep the carbon enhancement fixed at $\rm [C/Fe]=1,\ A(C)=7.56$ \citep[in agreement with the Galactic CEMP abundance trend, e.g. as reported by][Fig. 1]{Yoon2016}. Because the computation of 3D RHD stellar atmosphere models is computationally expensive, previous studies mostly relied on a single, representative set of stellar parameters \citep[e.g.][]{Gallagher2017,Collet2018}. To minimise uncertainties originating in the variety of stellar parameters in the data sample, we created solar-like MS models at $\rm T_{eff}=5750\ K,\ log(g)=4.5$, and cooler sub-giants at $\rm T_{eff}=5250\ K,\ log(g)=3.0$ for all chemical compositions. The chosen parameters represent a compromise between spanning the full range of stellar parameters and accurately reflecting the bulk of the distribution (see the discussion in Appendix \ref{appendix:model_parameter_selection} for further information). To ensure a satisfactory resolution, the number of grid points has been set to $240 \times 240 \times 120$. The physical size of the simulation domain has been limited between the average Rosseland optical depth of $\rm \log \tau_{ross} = -6$ and $+6.5$, which sets the geometric domain of the box. This is done to ensure that the models are deep enough to recover the adiabaticity at the bottom of the atmosphere and the relevant line formation region of the CH G band at the top. The resulting geometric domain ranges from approximately $2.5$ Mm for the MS 3D models (vertically) to 80 Mm for the sub-giant models. Horizontally, the models are twice as extended as vertically. \medis\ models are so-called box-in-a-star models, focusing only on the outermost layers of stars. Therefore, no additional assumptions about stellar parameters are necessary. For a typical mass of $\rm 0.8\, M_{\odot}$ \citep{Ruchti2013}, the MS model resembles a star with a radius of $\rm 0.83\, R_{\odot}$ and the sub-giant $\rm 4.7\, R_{\odot}$. A summary of the model parameters can be found in Table \ref{tab:models}.

For comparison, we used the classical 1D HE, scaled-solar, LTE model atmospheres from the MARCS grid \citep{Gustafsson2008}. To evaluate the effect of carbon enhancement also in 1D hydrostatic models, we furthermore obtained a carbon-enhanced MARCS model with stellar parameters $\rm T_{eff}=5750\ K,\ log(g)=4.5,\ [Fe/H]=-5, and\ [C/Fe]=+3.0$  to use in the analysis (B. Plez, priv. conv.).
\begin{figure}
    \centering
    \includegraphics[width=\columnwidth]{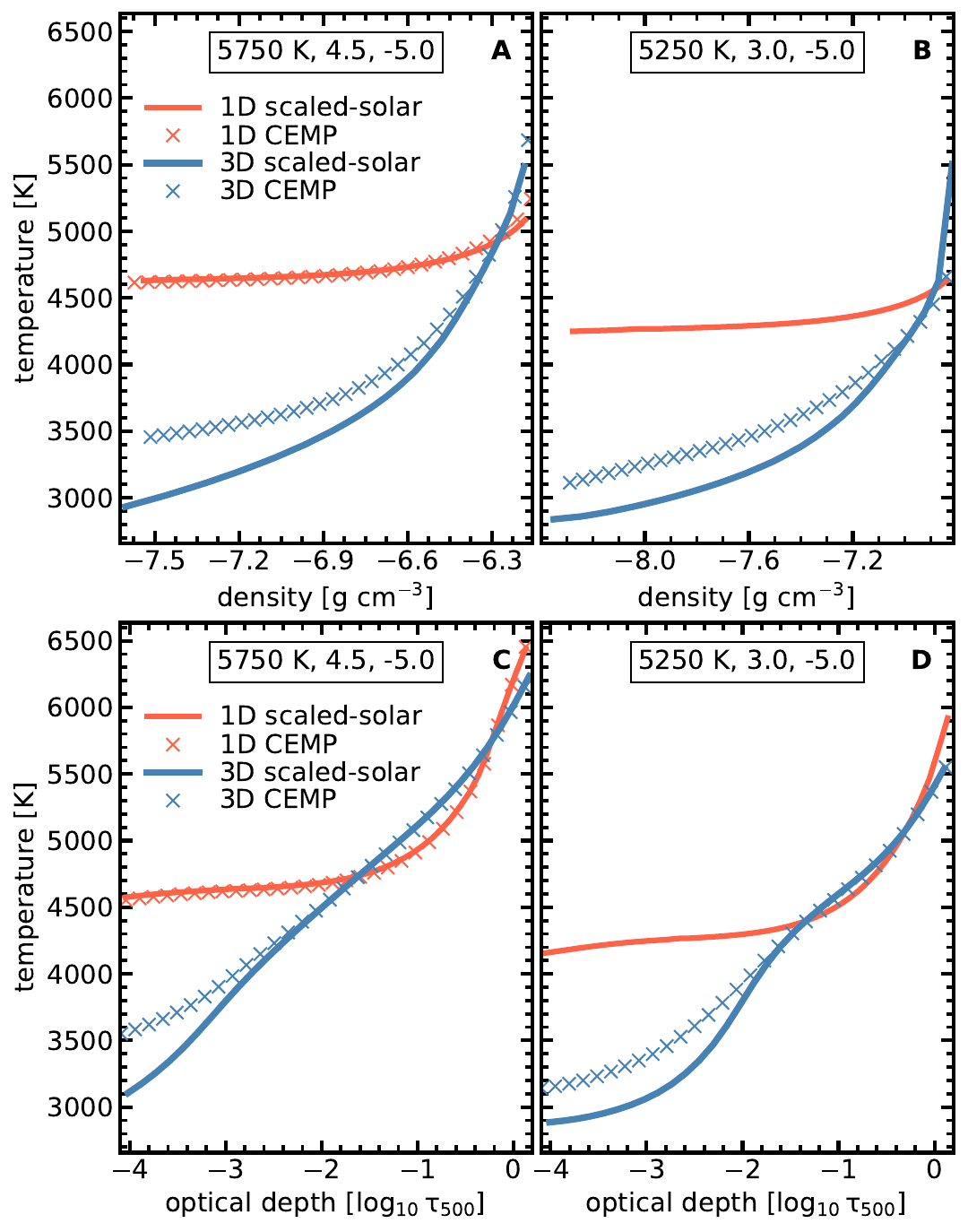}
    \caption{Comparison of the mean temperature stratification between 1D scaled-solar MARCS (solid red lines), 3D scaled-solar (solid blue lines), and 3D CEMP (blue crosses) models against mean density (top) and optical depth at $\rm 500\ nm$ (bottom). Left: MS model. We include a CEMP MARCS model (provided by B. Plez, priv. conv., shown with red crosses). Right: Sub-giant model. }
    \label{fig:ttau_ttho}
\end{figure}
\subsection{Spectrum synthesis}  \label{subsec:spectrum_synthesis}
\begin{figure*}
    \centering
    \includegraphics[width=1\columnwidth]{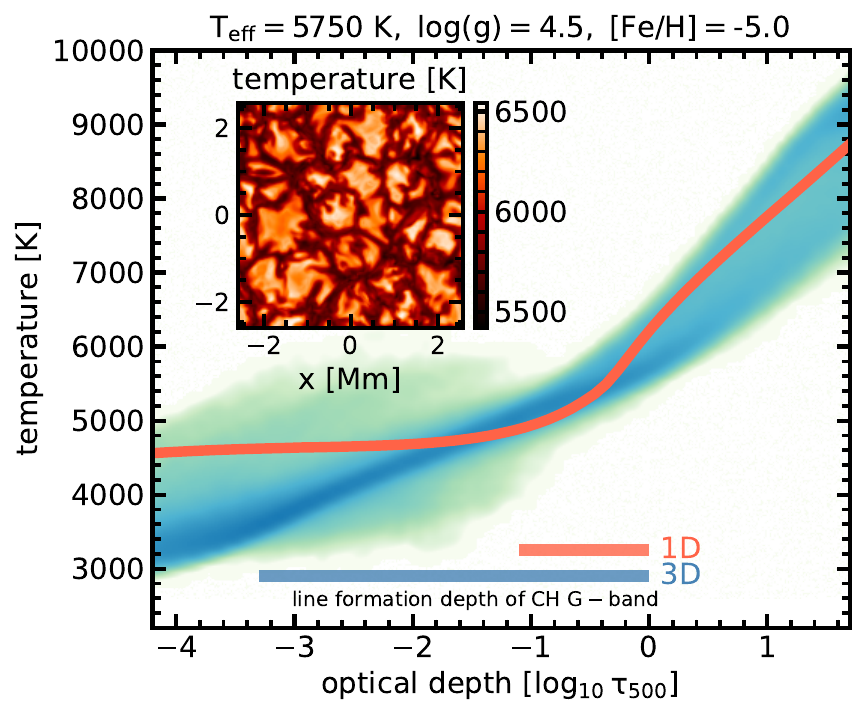}
    \includegraphics[width=1\columnwidth]{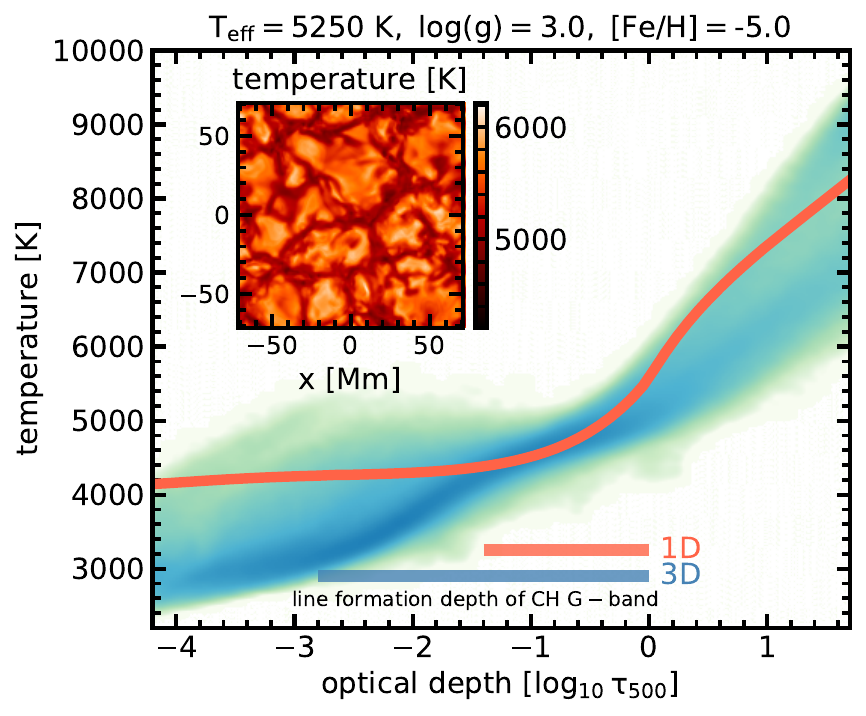}
    \caption{Comparison of the  mean temperature stratification between the \texttt{DISPATCH} 3D CEMP model and the 1D MARCS model on the optical depth scale. Left: $\rm T_{eff} = 5750\ K,\ log(g)=4.5$. Right: $\rm T_{eff} = 5250\ K,\ log(g)=3.0$. Horizontal bars at the bottom correspond to the approximate formation region of the G band in 1D and 3D.}
    \label{fig:average_profiles}
\end{figure*}

Synthetic spectra in 1D and 3D are computed using the updated version of \texttt{MULTI3D} code (we refer to \citealp{Eitner2024} for details). The code includes  extensive tables of bound-free and free-free radiative cross-sections for 92 chemical elements and partition functions from \cite{Irwin1981}. The occupation formalism is adopted for H I, following the HBOP routine by \cite{BarklemPiskunov2003}. The same code is used for the calculations with the MARCS and with \medis\ model atmospheres, to ensure the results are consistent.

We synthesised the CH G band in LTE between $\rm 4297\ and\ 4303\ \AA$ with a wavelength spacing of $0.001\ \AA$. In 3D, we down-sampled the full cube horizontally to a resolution of $\rm 20 \times 20$ points using linear interpolation. Such a horizontal down-sampling is sufficient to provide reliable profiles of spectral lines with abundance errors within $\sim \rm 0.01\ dex$ compared to simulations with a much higher horizontal resolution \citep{Bergemann2019, Bergemann2021, Rodriguez2024, Eitner2024}. The vertical resolution is not down-sampled in the spectrum synthesis, but rather up-sampled to an adaptive grid in the optically thin layers, to better resolve regions with large opacity gradients. For more information, see Hoppe et al. (in prep.). The final, re-sampled models have a resolution of $20\times20\times180$ points.

\section{Results}  \label{sec:results}
\subsection{3D model atmospheres}\label{subsec:cemp_atmospheres}
We begin with a qualitative discussion of the properties of 3D scaled-solar and 3D carbon-enhanced RHD model atmospheres in comparison to their 1D HE counterparts.

\subsubsection{3D scaled-solar versus 1D HE} \label{subsubsec:3Deffect}

First we compared the scaled-solar 3D RHD and 1D HE model atmospheres, computed using identical stellar parameters ($\rm T_{eff}$, $\rm \log g$, and metallicity). For simplicity, and to highlight the main physical differences, we discuss spatially averaged (on planes of constant optical depth) temperature-density (T-$\rm \rho$) profiles in this section but we emphasise that all subsequent spectrum synthesis calculations are performed in full 3D using 3D cubes. In Fig. \ref{fig:ttau_ttho} we show the T-$\rho$ (top) and T-$\tau$ (bottom) structures, where $\tau$ is the optical depth, of the MS (left, panels A and C) and of the sub-giant (right, panels B and D). As a representative case, we show both models at $\rm [Fe/H]=-5$ to investigate their properties and to illustrate the typical behaviour observed across the full set of models. All panels illustrate the respective atmosphere between $-4$ and $0$ in $\rm \log \tau_{500}$.

Comparing the 3D structures with their 1D counterparts, the dominant differences are apparent towards the upper layers of the atmospheres. For the MS model (Fig. \ref{fig:ttau_ttho}, panels A and C), the mean 3D temperature structure starts to deviate from that of the 1D HE model around optical depth $\rm \log \tau_{500}\sim-2$. The $\rm 3D - 1D$ difference reaches values of $\rm \sim -1\,500\ K$ at $\rm \log \tau_{500}=-4$. In those layers, the 3D RHD model is significantly and systematically cooler than the 1D MARCS model, which is well known and was extensively discussed in the earlier work on 3D models of very metal-poor MS and red giant stars \citep{Gallagher2016,Gallagher2017,Collet2018}. This difference is due to the fundamental property of convection, a process that is inherent to 3D RHD models, but is missing or parameterised by the mixing-length theory \citep{Gustafsson2008} in 1D HE models. In metal-poor stars, the atmospheric structure is dominated by adiabatic cooling of rising, and hence expanding, fluid parcels of the gas. This is in stark contrast to stars with solar-like metallicity: in such conditions, the high metallicity leads to significant radiative heating \citep{Stein1998}, which counteracts the adiabatic cooling effect and implies that the average 3D structure remains nearly in radiative equilibrium \cite[e.g.][]{Nordlund2009,Freytag2012,Magic2013a,Magic2013b}. 

A very similar behaviour can be seen in the structure of the 3D model of the sub-giant star (Fig. \ref{fig:ttau_ttho}, panels B and D). The 3D RHD model is over $\rm \sim -1\,000\ K$ cooler compared to its MARCS counterpart in the layers with $\log \tau_{ 500} \lesssim -3$. Although the effect is very large, it is slightly less than that of the MS 3D model in the uppermost layers ($\log \tau_{ 500} \sim -4$). One would naively expect that a sub-giant 3D model, due to lower pressures, would show larger deviations from the 1D HE model compared to the MS case. However, there is another interesting effect. The lower density and cooler structure of this model favours formation of different molecules, including CH, OH, and CO. Due to radiative re-heating, the temperature balance of the 3D model hence is slightly closer to that of 1D HE MARCS model. We emphasise that this difference is not a direct consequence of the C-enhancement, it is also present in models with scaled-solar composition, but of the overall absence of metal absorbers in the energy budget of the atmosphere.

\subsubsection{3D CEMP versus 3D scaled-solar} \label{subsubsec:Ceffect}
Comparing C-enhanced and scaled-solar 3D models directly (Fig. \ref{fig:ttau_ttho}, blue crosses and solid blue lines, respectively), we find that for the very metal-poor MS star (panels A and C), the effect is modest compared to the $\rm 3D - 1D$ differences described above. At optical depth of $\log \tau_{500}>-3$, there is no significant difference, while towards the upper boundary of the atmosphere, the mean temperature of the CEMP model is $\rm \sim 400 \ K$ hotter at $\rm \log \tau_{500}\sim -4$. In the sub-giant model (panels B and D), the same effect of C-enhancement is noticeable already at $\rm \log \tau_{500}\sim -2$, with temperature differences of $\rm \sim 200-300 \ K$ throughout the entire optically thin part of the atmosphere. Overall, the effect on the MS model is $\rm \sim 100\ K$ larger at around $\rm log\ \tau_{500}\lesssim -4$ than on the sub-giant. Towards deeper layers, however, $\rm \log \tau_{500}\gtrsim -3$, the scaled-solar sub-giant remains consistently cooler than its CEMP counterpart, while the departures between the MS models are already negligible.

The reason for this additional heating in the CEMP models is found in the increased abundance of C, which implies increased CH, CO, and CN molecular abundances, a higher molecular line opacity, and consequently more radiative heating of the upper atmospheric layers. Because O -- as well as N -- remain at their scaled-solar values, specifically the CH G band is one of the main contributors to the line opacity beyond $\rm \log \tau_{500}\sim -2$. In cooler atmospheres with very low [Fe/H], this effect is most noticeable. First of all, this is because already in the scaled-solar approximation C is an order of magnitude more abundant than Fe. In addition, the CH number densities are higher in cooler systems, as the chemical equilibrium favours a higher $\rm CH/C$ for lower T values.
The re-heating of the upper atmospheric layers induced by the C-enhancement hence produces the opposite of the $\rm 3D-1D$ effect described in Sect. \ref{subsubsec:3Deffect}. We furthermore note that there is no such effect in 1D HE MARCS models. To highlight this, we include one example CEMP MARCS model with the same stellar parameters in panels A and C.

Because of its lower $\rm log(g)$, the sub-giant model overall has a lower density at the same temperature as the MS model. A consequence of this is that the formation height of spectral lines generally shifts towards deeper layers in terms of optical depth. This explains the larger departures ($\rm 3D\ CEMP - \text{scaled-solar}$ models) at optical depth $\rm \log \tau_{500}\gtrsim-3$ among the sub-giant models compared to the MS models (compare the difference between blue lines and blue crosses in panel C with panel D) through the shift of the molecular line heating contribution to deeper atmospheric layers. Generally, decreasing the overall effective temperature causes the opposite effect, as the increased molecular number densities increase the opacity and hence shift the contribution function to lower optical depth (but see also the discussion in Appendix \ref{appendix:model_parameter_selection}).

\begin{figure}
    \centering
    \includegraphics[width=1\columnwidth]{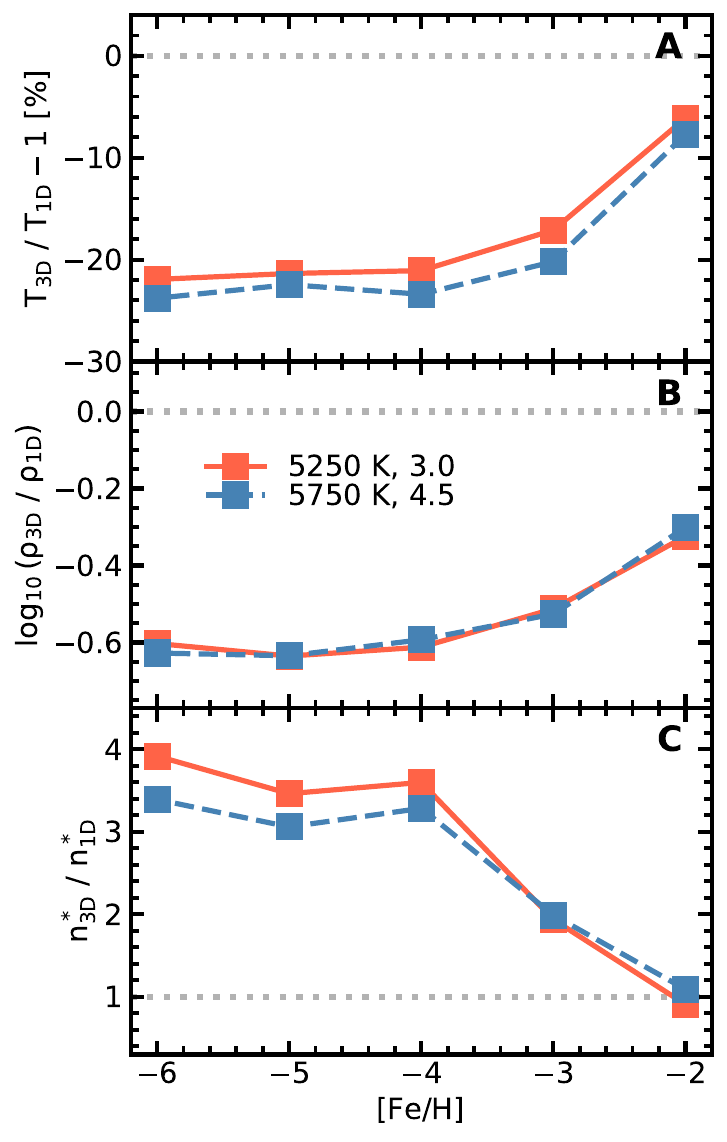}
    \caption{Difference between 3D CEMP and 1D MARCS models as a function of metallicity. Top: Relative temperature difference (in percent). Middle: Logarithmic density differences (in dex). Bottom: CH/H number density fraction, where $\rm n^*=n_{CH}/n_{H}$ is the CH number density relative to the hydrogen number density. Dashed blue lines correspond to $\rm T_{eff} = 5750\ K,\ log(g)=4.5$ models, solid red curves to $\rm T_{eff} = 5250\ K,\ log(g)=3.0$. The differences in each figure are computed at the representative formation height of the strong lines in the G band. See Sect. \ref{subsec:spectra} for more information.}
    \label{fig:feh_diff_1D}
\end{figure}
\begin{figure*}
    \centering
    \includegraphics[width=1\textwidth]{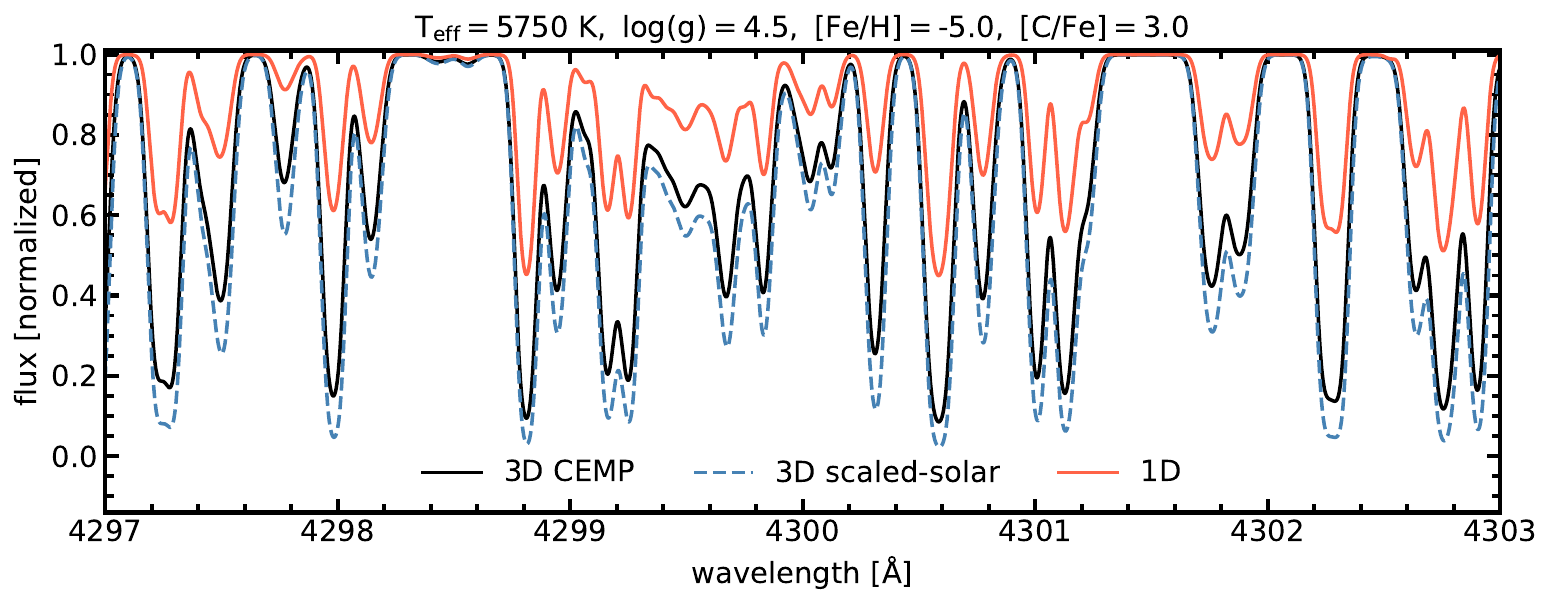}
    \includegraphics[width=1\textwidth]{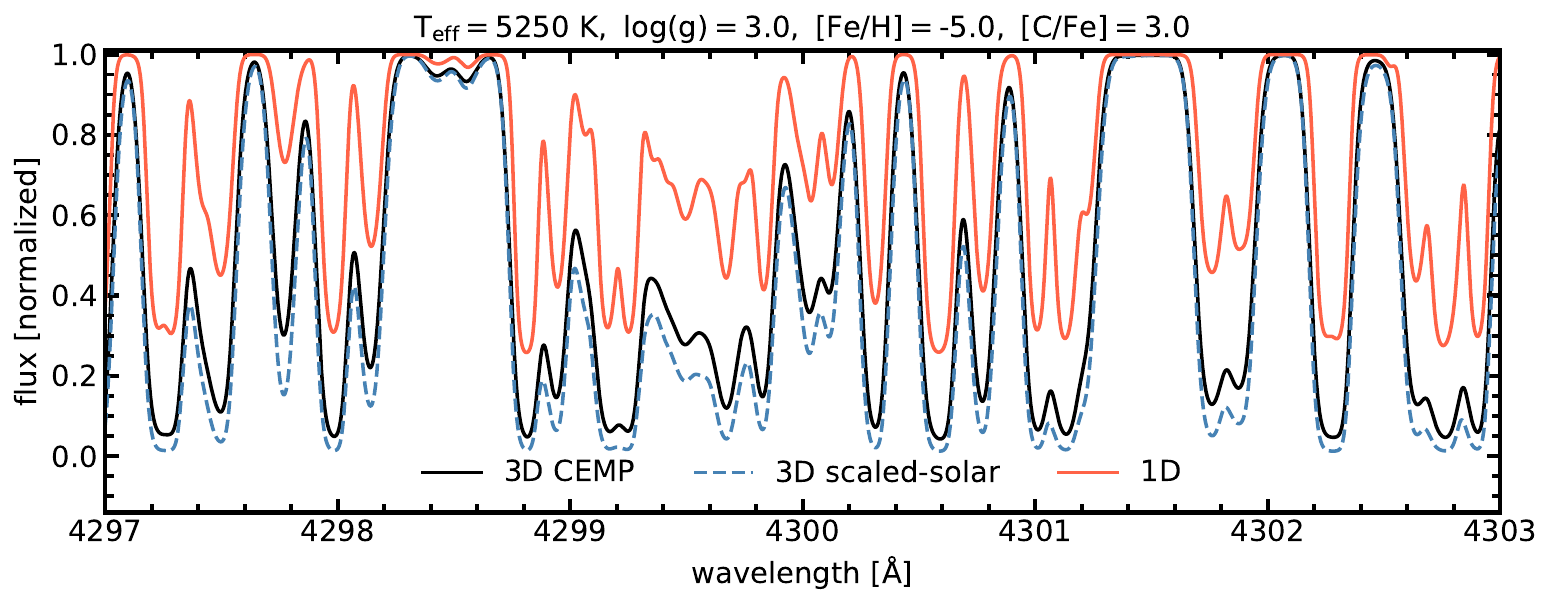}
    \caption{CH G band computed with \texttt{M3DIS} 3D and 1D MARCS model atmospheres. Solid black lines represent the CEMP 3D model, dashed blue lines the corresponding scaled-solar 3D model. Solid red lines show the 1D MARCS model with the same stellar parameters. Top: $\rm T_{eff} = 5750\ K,\ log(g)=4.5$. Bottom: $\rm T_{eff} = 5250\ K,\ log(g)=3.0$.}
    \label{fig:spectra}
\end{figure*}
\subsubsection{3D CEMP versus 1D HE} \label{subsubsec:3dCEMP1d}
Combining the results from Sects. \ref{subsubsec:3Deffect} and \ref{subsubsec:Ceffect}, we next compared our \medis\ 3D CEMP model atmospheres with 1D HE scaled-solar MARCS models. The difference between these two model types is particularly important, as many studies rely on 1D scaled-solar MARCS models for spectral synthesis and abundance analysis. This raises the question of whether this simplifying assumption introduces systematic effects in the derived abundances.

To find the answer, we first show the full 3D temperature structure of both selected 3D CEMP models at $\rm [Fe/H]=-5$ in Fig. \ref{fig:average_profiles}. For both models, MS (left-hand side) and sub-giant (right-hand side), we used the C-enhancement of $\rm[C/Fe] =+3$. We furthermore show in the inset a horizontal slice of the same atmospheres, interpolated to the optical surface at $\rm \log \tau_{500}=0$. We note the very different scales and granular size between the MS (left panel) and the sub-giant (right panel) models. The model of the MS star is, because of a much stronger gravity, about 20 times smaller than that of a sub-giant simulation, the latter exceeding 100 Mm in size. As described in the previous sections, the difference between 3D CEMP and 1D MARCS models is most striking in the optically thin layers of the atmosphere, where 3D CEMP models are still significantly cooler, even though the difference is reduced by the C-enhancement-induced re-heating. The two models converge towards the optical surface; the agreement is best in the proximity of $\rm \tau=2/3$, indicating a very similar effective temperature of 1D and 3D models.

Towards the bottom of Fig. \ref{fig:average_profiles} we furthermore include horizontal bars, indicating the formation region of the G band. For this, we selected a representative CH line, compute the contribution function in its core for each model atmosphere, and used the location of its peak as proxy for the optical depth at which the G band typically forms. We drew the horizontal lines until $\rm \tau_{500}=1$, which corresponds to the formation of weaker features, line wings and the continuum. We emphasise that the computation of the G-band formation region does not affect the results or conclusions of this study; it is intended solely as a useful reference to aid in the analysis of model differences and to support their physical interpretation. In Fig. \ref{fig:average_profiles} we see that the G band forms slightly deeper within the atmosphere of the sub-giant as compared to the MS model, which is a consequence of the significantly lower surface gravity and hence overall lower density. This shift happens despite the lower effective temperature of the 3D sub-giant model. In the 1D model the upper layers are strongly dominated by radiative equilibrium -- and hence the effective temperature -- which means the drop in temperature between the MS and sub-giant models is more significant, the opacity rises and the formation region of the G band shifts in the opposite direction.

The evolution of the differences between 3D CEMP and 1D MARCS models with metallicity can be investigated in Fig. \ref{fig:feh_diff_1D}. For clarity, we refrain from showing the differences between the full profiles as a function of optical depth for each metallicity. Instead, we present a representative 3D–1D difference for each metallicity, computed from the atmospheric regions most relevant to G-band formation. Accordingly, the data shown in this figure are evaluated at the G-band formation height in each respective atmosphere, as described in the previous paragraph.

In panel A of Fig. \ref{fig:feh_diff_1D} the relative temperature difference is shown for the sub-giant models (solid red lines) and for the MS models (dashed blue lines). We find that 3D CEMP models are on average between $20\%$ and $25\%$ cooler than 1D HE models. At metallicities $-6$, $-5$, and $-4$, there is no significant variation in the 3D–1D difference. At $\rm [Fe/H]=-3$ the differences start to decline, and reach values around $5\%$ at metallicity $-2$. Besides forming in cooler parts of the atmosphere, the G band furthermore forms in regions of lower densities (panel B), with $\rm \log \rho_{3D}/\rho_{1D}$ between $-0.6$ dex at $\rm [Fe/H]=-6$ and $-0.3$ dex at $\rm [Fe/H]=-2$. Note that this is true despite the fact that metal-poor 1D HE models generally are less dense at the same optical depth (due to the hotter temperatures) and the data in panel B mostly reflect the differences in formation height through the exponential growth of density with depth.

Additionally, the MS models show 3D CEMP effects similar to -- if not even slightly larger than -- the sub-giant models, especially when considering the temperature differences (Fig. \ref{fig:feh_diff_1D}, panel A). However, at this stage it is important to consider that the molecular equilibrium, which ultimately determines the CH number densities and hence directly influences the strength of the spectral lines in the G band, is not a linear function of temperature and density, but rather exhibits a temperature and density dependence as dictated by the Saha equation \citep[e.g.][]{Carson1992}:
\begin{equation} \label{eq:moleq}\rm
    \frac{n_{C}n_H}{n_{CH}} \propto \frac{z_C z_H}{z_{CH}}\ T^{3/2}\ e^{\frac{-D_{CH}}{k_BT}}\ ,
\end{equation}
where $\rm n_x$ is the number density of species $\rm x$, $\rm z_x$ the partition functions and $\rm D_{CH}$ the dissociation energy of the CH molecule, which is set to $\rm 3.47\ eV$ \citep[][and references therein]{Popa2023,Storm2025}. To highlight this non-linear dependence, we include panel C, which shows the fraction of CH/H number densities $\rm n^*_{3D}/n^*_{1D}$ -- where $\rm n^* = n_{CH}/n_{H}$ -- in the same formation region as panels A and B. Here it becomes obvious that the number density of CH/H molecules in the sub-giants deviate by a factor of $\rm \sim 1.2$ more between 3D and 1D than what is found in MS models. So despite showing a slightly smaller 3D effect in temperature and density, the 3D effect is more prominent in the CH/H number density of sub-giants, which will ultimately be reflected in the G band and $\rm [C/Fe]$ abundance corrections (see Sect. \ref{sec:conclusions}).

Overall, with increasing [Fe/H], the differences in the structure between 3D CEMP and 1D HE models decrease noticeably. The reason is that with increasing metal content, radiative heating starts to dominate the overall energy budget of the photosphere. The consequence is that adiabatic cooling becomes less dominant in the outer layers, which ultimately closes the gap between the average 3D structure and 1D HE MARCS models, which do not include this cooling mechanism in the first place.

Our results are in a good agreement with work presented by \cite{Gallagher2016,Gallagher2017}, although a different code \citep[\texttt{CO5BOLD}:][]{Freytag2012} was used in that study and their choice of stellar parameters does not fully overlap with that of our models. \cite{Gallagher2017} furthermore note that if the O abundance is changed, this has an impact on the number densities of the CO molecule. Consequently, the amount of free C available to form CH is decreased. Whereas generally it would be of interest to test non-solar abundances of different elements, in this work we did not explore the dependence on the $\rm C/O$ ratio. First, because there is evidence of CEMP stars with only very modest O enhancement \citep{Bessel2004,Iwamoto2005}. Also, most diagnostic OH features that are suitable for O abundance measurements \citep{Asplund2001} are located in the UV range, which is extremely sensitive to non-LTE \revisedPE{(NLTE)} effects in metal-poor stars \citep{Bergemann2014,Lind2024}. Furthermore, no 3D \revisedPE{NLTE} $\rm C/O$ ratios are available for CEMP stars \citep{Collet2018}.
We also note that \cite{Gallagher2017} focus on a hot sub-giant with metallicity $-3$, which leads to smaller differences between 1D and 3D models -- partly because of the higher metallicity, and partly because of the lower $\rm CH/C$ ratio in both CEMP and non-CEMP models. We explore the impact of different metallicities and stellar parameters on synthetic stellar spectra and C abundance determinations in the following sections.

\subsection{Spectrum synthesis of the CH G band}\label{subsec:spectra}
We synthesised the CH G band for all models listed in Table \ref{tab:models}, including their scaled-solar counterpart, and for comparison, the respective 1D MARCS models with the same stellar parameters. 

In Fig. \ref{fig:spectra} we show two representative synthetic G-band profiles for two model atmospheres, MS (top) and sub-giant (bottom), at $\rm [Fe/H]=-5$ and $\rm [C/Fe]=+3$. Independent of stellar parameters, it is evident that the 3D CEMP model (solid black lines) leads to significantly stronger CH features than the 1D model (solid red lines) at the same $\rm A(C)$. The 3D scaled-solar model (dashed blue lines) on the other hand produces an even stronger G band as the 3D CEMP model (for the same T$_{\rm{eff}}$, $\rm \log(g)$, and metallicity) and hence shows the largest departure from 1D HE. This is reflected in the equivalent width (EW).

\begin{figure}
    \centering
    \includegraphics[width=1\columnwidth]{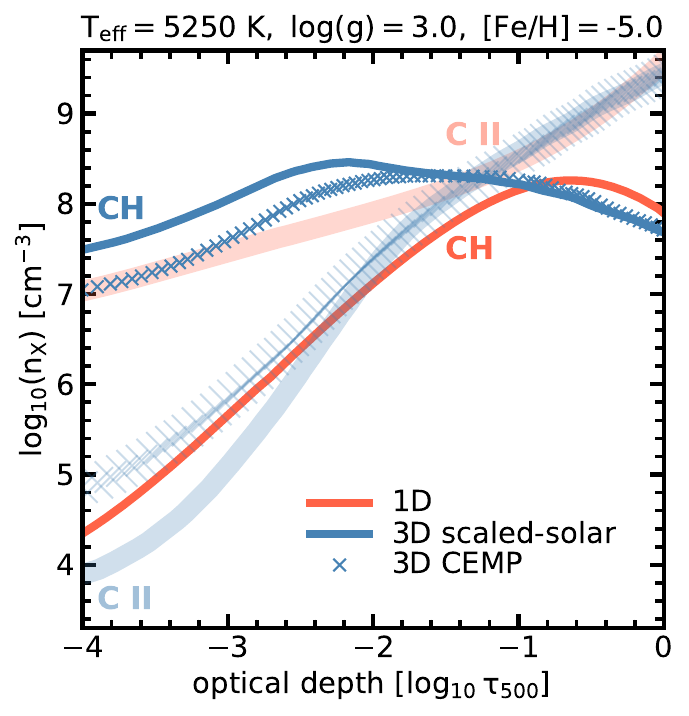}
    \caption{CH and C \uproman{2} number densities as a function of optical depth. Solid blue lines correspond to the 3D scaled-solar model, blue crosses to the 3D CEMP model. The 1D MARCS model is shown with solid red lines. For all models, CH number densities are shown as opaque lines, while C \uproman{2} number densities are shown as thick, partially transparent lines.}
    \label{fig:molec_abund}
\end{figure}
\begin{figure*}
    \centering
    \includegraphics[width=\textwidth]{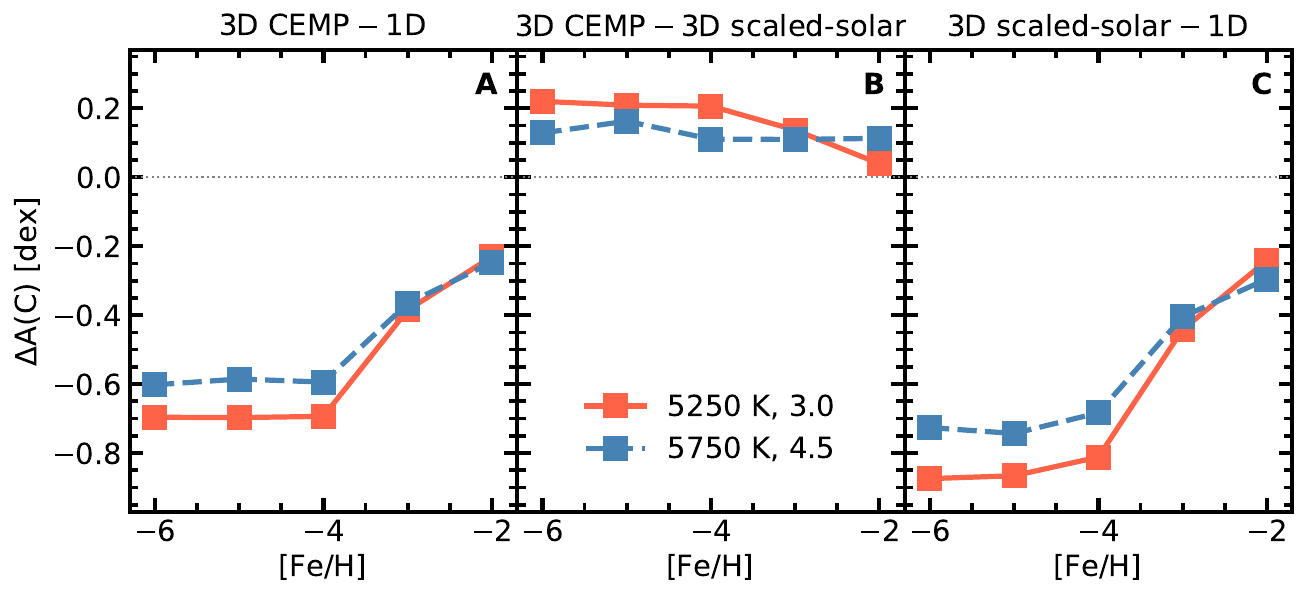}
    \caption{Abundance corrections to A(C) derived from the EW of the CH G band between $\rm 4\,297\ and\ 4\,303 \AA$. Panel A: Final abundance correction $\rm \Delta A(C)=A(C)_{3D\ CEMP}-A(C)_{1D}$ from $\rm 1D\ HE\ MARCS$ to $\rm 3D\ CEMP$ models. Panel B: $\rm A(C)_{3D\ CEMP} - A(C)_{3D\ \text{scaled-solar}}$. Panel C: $\rm A(C)_{3D\ \text{scaled-solar}} - A(C)_{1D}$.}
    \label{fig:corrections}
\end{figure*}

These strong differences in the G-band strength are due to the differences in the atmospheric structure. To explore this connection in more detail, we show the average CH number density, $\rm \log n_{CH}$, in the atmosphere of the sub-giant model at $\rm [Fe/H]=-5$ as a function of optical depth in Fig. \ref{fig:molec_abund}. We additionally include the density of singly ionised carbon (C \uproman{2}). The CH number density can best be understood by linking their evolution with optical depth to the differences in the atmospheric structures as presented in Fig. \ref{fig:ttau_ttho}, panel D. Below $\rm \log \tau_{500}=-1$ the atmosphere is dropping sharply in temperature -- in 3D as well as in 1D -- which favours the formation of CH molecules at the expense of C \uproman{2}, which is recombining with free electrons in the successively cooler and less dense environment. Towards the upper boundary the temperature in the 1D HE model stays roughly constant, while the density is dropping exponentially. In chemical equilibrium the CH number density is proportional to $\rm n_{C}$ -- as well as $\rm n_{H}$ (Eq. \ref{eq:moleq}) -- and hence $\rm n_{CH}$ drops proportional to $\rm \rho^{2}$. This explains why the CH number density decreases significantly in the 1D model, despite the temperature staying constant. The C \uproman{2} number density on the other hand is only linearly dependent on density, and additionally is affected by the balance of $\rm H^-$ and C \uproman{2} dissociation energies, which is why the trend is less obvious.

While this proportionality is also true for the 3D model, the behaviour in the atmosphere is entirely different. This is because the 3D model is not in radiative equilibrium -- it is dominated by adiabatic cooling -- which means the temperature drops as a function of decreasing optical depth. In an increasingly colder environment the conditions for CH formation improve drastically, which is why $\rm n_{CH}$ increases in both 3D models and departures between 3D and 1D become significant. Comparing 3D CEMP (blue crosses) and 3D scaled-solar (solid blue lines), we see that there are significantly more CH molecules present in the upper layers of the scaled-solar model, while there is only very little C \uproman{2}, compared to the CEMP model. Looking again at Fig. \ref{fig:ttau_ttho} this difference can be attributed to the additional heating induced by the C-enhancement in the CEMP atmosphere, which causes it to be hotter and hence less favourable for CH formation.

3D RHD models are much, up to $-1000$ K, colder in atmospheric layers above $\log \tau_{500} \lesssim -2.5$. Because of the strong C-enhancement, the cores of strongest lines in the G band form in regions around $\rm \log \tau_{500}=-3.5$, as indicated by the horizontal lines in Fig. \ref{fig:average_profiles} and the $\rm n_{CH}$ extrema in Fig. \ref{fig:molec_abund}. Due to much lower temperatures, the formation of molecules is more efficient, and due to larger molecular number densities the absorption profiles are stronger \citep{Uitenbroek2011}. The final consequence is that the presence of strong G-band absorption in an observed spectrum is not necessarily linked to a significant C-enhancement but can, in part, be associated with strong convection as expected at low metallicities \citep{Collet2011,Gallagher2017}.

The larger differences seen in the synthetic spectra for the sub-giant (Fig. \ref{fig:spectra}, bottom panel), as compared to the MS star, directly reflect the atmospheric structure distinctions discussed in Sect. \ref{subsubsec:3dCEMP1d}. Specifically, while the 1D sub-giant model is colder (enhancing its G band), the interplay of the non-linear dependence of CH formation on temperature and density, combined with the more pronounced C-enhancement induced re-heating effect in the 3D sub-giant atmosphere, leads to the observed spectral differences relative to 1D.

\subsection{CEMP abundances in 3D}\label{subsec:corrections}
In order to estimate the influence of carbon-enhanced CEMP 3D RHD models on synthetic observables, we performed the analysis as presented in the previous sections for stellar model atmospheres of different metallicities, from $\rm [Fe/H]=-2$ up to $\rm [Fe/H]=-6$. We compared the total EWs in the wavelength range of the G band obtained in 3D with the corresponding 1D MARCS results, computed using a micro-turbulence of 1 km$/$s. By varying A(C) in the spectrum synthesis of the 3D RHD models, we sampled the full curve of growth and derived abundance corrections by matching the EWs within the integration limit of 4297 to 4303 \AA. We emphasise that this EW-integration of the G band is not used to obtain abundances of observed spectra directly, but only to quantify the corresponding overall effect on carbon abundances by comparing synthetic spectra. Other procedures, such as a direct fitting of lines \citep{Popa2023}, can be used instead.

Our results for the abundance corrections can be found in Fig. \ref{fig:corrections}. We present the final correction 
$\rm \Delta A(C) = A(C)_{3D\ CEMP}-A(C)_{1D}$ in panel A, the CEMP correction $\rm A(C)_{3D\ CEMP}-A(C)_{3D\ \text{scaled-solar}}$ in panel B, and the $\rm A(C)_{3D\ \text{scaled-solar}} - A(C)_{1D}$ correction in panel C for the MS and sub-giant model as a function of metallicity. We note that, because we compare $\rm 3D\ CEMP$ to the $\rm 1D\ \text{scaled-solar}\ MARCS$ models, in panel A the effect of carbon enhancement on the correction is implicitly taken into account. We opted to use scaled-solar rather than C-enhanced MARCS models for our comparison, since scaled-solar models are more commonly employed in 1D studies of CEMP stars. However, panels A and C of Fig. \ref{fig:ttau_ttho} indicate that the differences between the two model types are negligible, making this distinction ultimately unimportant.

As explained in the previous section, $\rm 3D-1D$ abundance corrections are larger (in modulus, or amplitude) for the sub-giant star. At $\rm [Fe/H]=-6$ they are large and negative, and reach up to $\rm -0.7\ dex$ with C-enhancement in the atmosphere, and $\rm -0.9\ dex$ without. Corrections decrease with increasing metallicity and are smaller, $\rm\Delta A(C) \approx -0.2 \ dex$, at $\rm [Fe/H]=-2$. The $\rm 3D\ CEMP - 1D$ corrections for the hotter MS star range from $\rm \sim-0.6\ dex$ at [Fe/H] $=-6$ to $\rm \sim -0.2\ dex$ at [Fe/H] $=-2$. 
In contrast, the effect of carbon enhancement in 3D models is positive (panel B). At [Fe/H] $=-2$, 3D scaled-solar and 3D CEMP G-band strengths in the sub-giant models are almost identical, which is why $\rm A(C)_{3D\ CEMP}-A(C)_{3D\ \text{scaled-solar}} \lesssim 0.05\ dex$. The differences increase with decreasing metallicity and decreasing surface gravity, so that $\rm A(C)_{3D\ CEMP}-A(C)_{3D\ \text{scaled-solar}} = \rm +0.2\ dex$ for the sub-giant model with $\rm [Fe/H]=-6$. As discussed in Sect. \ref{subsubsec:Ceffect}, the effect of C-enhancement on the hotter MS model is overall smaller. The corrections are of the order of $\rm \lesssim +0.1\ dex$, and show only a small deviation with metallicity. Compared to the significant 3D effect, the C-enhancement in hot, low metallicity MS models is hence negligible.
\begin{figure}
    \centering
    \includegraphics[width=\columnwidth]{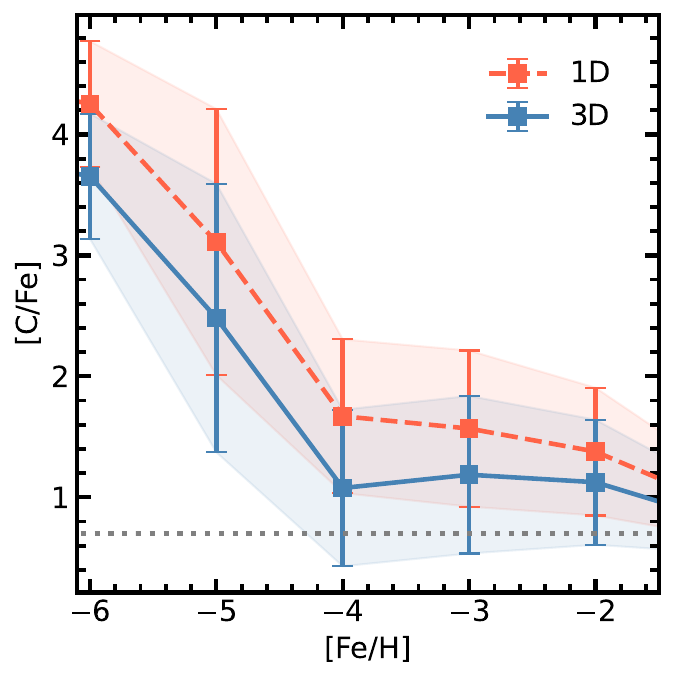}
    \caption{[C/Fe] against [Fe/H] for 3D CEMP and 1D measurements of Galactic stars from a MS + sub-giant subsample of the SAGA database. The dashed red line represents the mean, uncorrected 1D and the solid blue line the 3D CEMP-corrected data. Shaded regions correspond to standard deviations of the observed sample in the respective metallicity bin $\rm \pm 0.5\ dex$. The dotted grey line represents the dividing CEMP classification abundance of $\rm [C/Fe]=0.7$.}
    \label{fig:cfe}
\end{figure}
\begin{figure}
    \centering
    \includegraphics[width=\columnwidth]{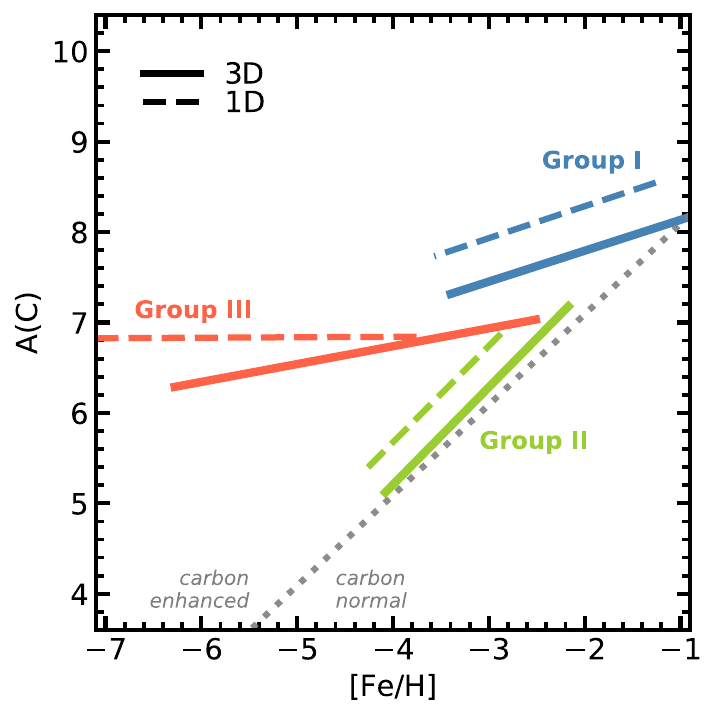}
    \caption{Yoon-Beers \citep{Yoon2016} diagram of A(C) against [Fe/H] for the 3D-corrected measurements for Galactic stars from the literature. The dotted grey line represents the dividing CEMP classification abundance of $\rm [C/Fe]=0.7$. Coloured lines show the major axis of the ellipse representing the distribution of different CEMP groups. In 1D (dashed lines), ellipses are drawn similar to \cite{Yoon2016}, their Fig. 1. In 3D (solid lines) ellipses are obtained from fitting a Gaussian mixture model to the data.}
    \label{fig:yoon-beers}
\end{figure}

It is interesting to analyse the impact of 3D CEMP models in the context of the observed CEMP distribution of Galactic stars. We therefore applied our corrections to a sample of MS and turnoff stars compiled from the SAGA database \citep{Suda2008,Suda2011,Suda2017} to explore whether 3D models have an impact on statistics of CEMP stars in different Galactic metallicity regimes. We applied $\rm A(C)_{3D\ CEMP} - A(C)_{1D}$ carbon abundance corrections to the observed data, based on the closest 3D model in terms of $\rm T_{eff}$ and $\rm log(g)$, and we interpolated linearly in metallicity. We then split the stellar sample into bins of $\rm \pm 0.5\ dex$ in metallicity and computed the mean and standard deviation ($\sigma$) in each of them. The result is shown in Fig. \ref{fig:cfe}. As expected, the differences between the 3D-corrected and 1D  distributions increase with decreasing [Fe/H]. At $\rm [Fe/H]=-4$ the mean $\rm [C/Fe]$ of CEMP stars is $\rm \approx 1.1$, $\rm 0.5\ dex$ lower than the $\rm [C/Fe] \approx 1.6$ that the 1D data suggest. Towards $\rm [Fe/H]=-2$ the differences decrease to $\rm \approx 0.2 \ dex$. We note that at metallicities above $-5$ the lower limit of the 1-$\sigma$ interval already lies below the CEMP defining line of $\rm [C/Fe]=0.7$ (indicated by the dotted grey line) in 3D, which implies that there is a significant impact on the overall abundance distribution in this metallicity regime. Below $\rm [Fe/H]=-5$ the corrections are larger; however, because of the overall higher $\rm [C/Fe]$ abundances found in these stars, the impact on the distribution of CEMP stars is expected to be small.

To investigate Galactic CEMP populations further, we present the same data in the so-called Yoon-Beers diagram \citep{Yoon2016}, which is shown in Fig. \ref{fig:yoon-beers}. In this figure, we show the location of CEMP subgroups \uproman{1}, \uproman{2}, and \uproman{3} as defined by \cite{Yoon2016}; see their Fig. 1. In 1D, the ellipses are adopted from \cite{Yoon2016}, while in 3D we fitted a Gaussian mixture model to our data. In addition to the first identification of CEMP subgroups, \citeauthor{Yoon2016} furthermore showed that group \uproman{1} is mainly populated by CEMP-r/s stars, while groups \uproman{2} and \uproman{3} contain mainly CEMP-no stars. \cite{Yoon2016} -- and subsequently also \cite{Yoon2018,Chiaki2017} -- furthermore identify different CEMP groups with different formation scenarios, indicating multiple progenitors for CEMP-no stars.

At lower metallicities, differences manifest as a systematic offset of $\rm \sim 0.6\ dex$, shifting down the $\rm A(C)$ plateau (indicated by the horizontal orientation of the red line in the 1D case) significantly. We note that, as a consequence of this plateau, the carbon enhancement in the atmosphere, relative to iron, becomes more significant when considering even lower metallicities. We do not expect that -- at least in CEMP stars -- G bands from 3D and 1D models diverge even further at more extreme metallicities, mainly because the carbon enhancement itself will compensate for the loss of metals in exactly the right part of the atmosphere, as demonstrated in the previous sections. This compensation is noticeable as a decreasing metallicity gradient in panel A (compared to panel C) of Fig. \ref{fig:corrections} already between $\rm [Fe/H]=-4\ and -5$. One may be tempted to use scaled-solar, low metallicity, non-carbon-enhanced models for the analysis of these CEMP stars due to the time investment related to creating 3D models of stellar atmospheres. However, this would lead to an overestimation of the corrections and introduce an additional differential effect with metallicity that causes the A(C) plateau to be artificially tilted. When properly including C-enhancement in the atmosphere computation, this tilt is weaker, but still noticeable mostly due to differences in stellar parameters of individual targets. In Fig. \ref{fig:yoon-beers} we find that an ellipse tilted by $\rm 10^{\circ}$ with respect to the 1D case best represents our group \uproman{3} sample in 3D.

Because $\rm 3D-1D$ abundance corrections are negative, we find a significant number of stars in the SAGA sample that are -- following the classical definition of \cite{Aoki2006} -- no longer CEMP stars, as their $\rm [C/Fe]$ drops below $\rm 0.7$. This is most noticeable in CEMP group \uproman{2}, as these stars are located closest to the carbon-normal division in terms of $\rm A(C)$, while still experiencing a large abundance correction due to their relatively low metallicities.
Around the onset of the carbon-plateau at $\rm [Fe/H]=-4$ we find that stars at higher $\rm [C/Fe]$ are systematically redistributed towards the abundance threshold, such that close to -- and below -- the dividing line we find significantly more stars in 3D than in 1D. Overall, there are significantly fewer stars with extreme carbon abundances above $\rm [C/Fe] = 1.5$ (see Fig. \ref{fig:cfe}, specifically for $\rm [Fe/H] \geq -4$). The consequence is that that there is a migration of stars at the edges of groups \uproman{1} and \uproman{3} towards group \uproman{2}, or even outside the CEMP category. In 3D, we find that group \uproman{2} is best represented by an ellipse that is shifted down by $\rm 0.4\ dex$ with respect to the 1D case.

\begin{figure}
    \centering
    \includegraphics[width=\columnwidth]{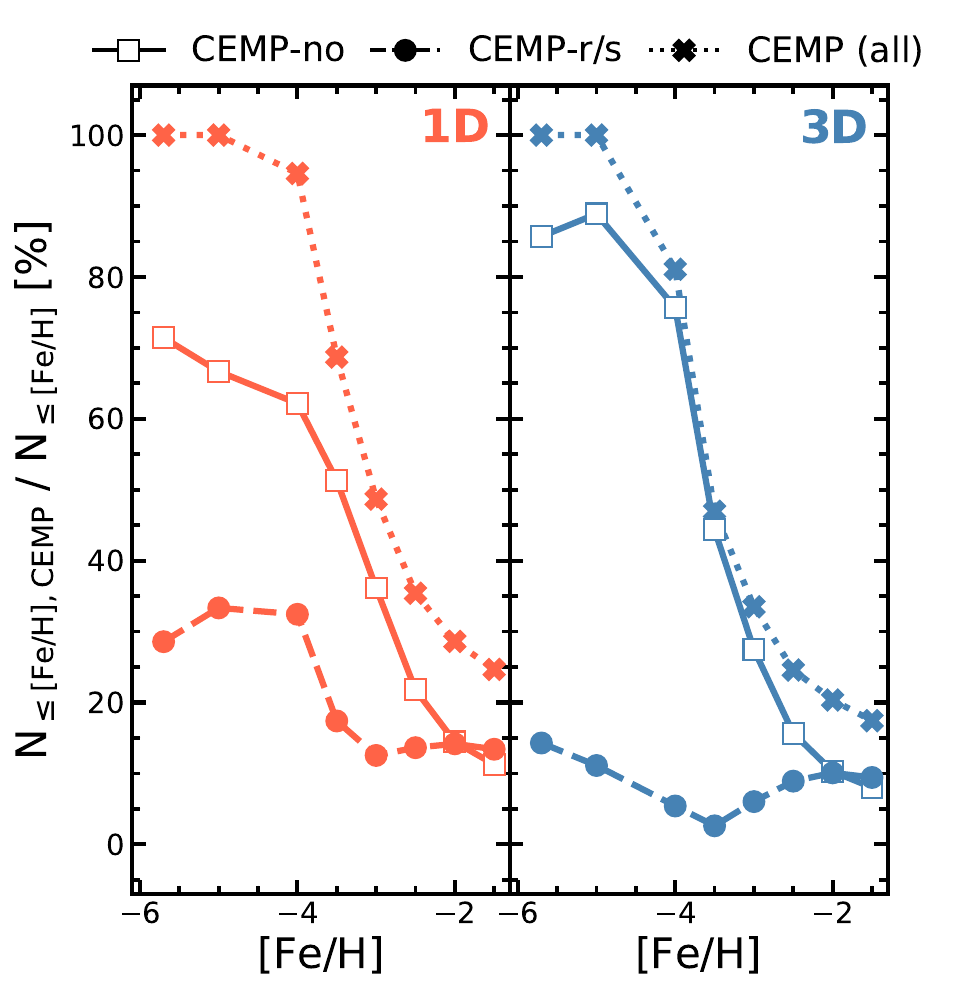}
    \caption{Cumulative fraction of CEMP-no (solid lines with squares) and CEMP-r/s (dashed lines with circles) stars in 1D (red, left panel) and 3D (blue, right panel). The dotted line with crosses shows the total cumulative fraction of CEMP stars in 1D (left) and 3D (right).}
    \label{fig:cumsum_rs_no}
\end{figure}

\subsection{Galactic CEMP fraction in 3D}\label{subsec:cemprs}

To quantify the consequences for Galactic fractions of CEMP-r/s and CEMP-no subgroups, splitting the CEMP sample based on the neutron capture abundances is necessary. However, for a significant portion of the SAGA sample no Ba or Eu measurements are available. This issue has been discussed by \cite{Yoon2016}, who derived an alternative classification based on $\rm A(C)$ alone, for which they claim a success rate of $\rm 87\%$ for CEMP-r/s classification and $\rm 93\%$ for CEMP-no stars. In their method, stars are associated with the CEMP-r/s group, if a high absolute carbon abundance $\rm A(C)>7.1$ is measured, while CEMP-no stars consequently have $\rm A(C)\leq 7.1$. The discussion of their approach goes beyond the boundaries of this paper. We emphasise that their CEMP subgroup association is empirical, which means the systematic offsets presented here will either lead to a re-classification of CEMP stars in the same system, if the separation of \cite{Yoon2016} is still valid, or require a change of the system through new $\rm A(C)$ limits based on new C abundances. 

In the following, we utilise their method to classify our sample twice, in 1D and after the 3D CEMP abundance corrections have been applied. The resulting cumulative fractions can be found in Fig. \ref{fig:cumsum_rs_no}. We show the cumulative fraction for CEMP-r/s (dashed lines) and CEMP-no (solid lines) stars separately -- as well as for their sum (dotted lines) -- for the 1D (red, left panel) and 3D (blue, right panel) case. 

In 1D, there is a significant fraction of over 10 $\%$ of CEMP-r/s stars across all metallicities; below $\rm [Fe/H]=-4$ we find $\sim 30 \%$ of all stars in our sample are CEMP-r/s stars. The fraction of CEMP-no stars has a strong metallicity dependence. Below $\rm [Fe/H]=-5$ all stars in the sample are CEMP stars, with a share of $70\%$ CEMP-no and $30\%$ CEMP-r/s. Increasing metallicity decreases the contribution from CEMP-no stars, until $\rm [Fe/H]\leq-2$, where the two contribute equally to the total number of CEMP stars ($\sim 15\%$ each).

However, in 3D, there are overall $\sim 10\%$ fewer CEMP stars below $\rm [Fe/H]=-2$. Furthermore, the fraction of CEMP-r/s stars is significantly lower, which is due to the increasingly negative (with decreasing [Fe/H]) 3D abundance corrections and the resulting re-distribution of group \uproman{1} and \uproman{3} stars -- where most of the CEMP-r/s stars reside -- to the CEMP-no dominated group \uproman{2}. CEMP-r/s stars are preferentially found at higher metallicities than CEMP-no stars, where the C-enhancement in $\rm [C/Fe]$ is less severe than at low metallicities; A large number of CEMP-r/s stars reside close to the defining boundary of $\rm [C/Fe]=0.7$. When 3D corrections to C abundances are taken into account, many stars originally classified as CEMP-r/s stars in reality have a lower C-abundance, implying their identity as CEMP-no stars. The contribution from CEMP-no stars hence increases with respect to 1D, and climbs to above $90\%$ for all stars below $\rm [Fe/H]=-5$, while it peaks around $70\%$ in 1D.

We remind the reader that it must still be explored, using large self-consistent stellar samples with 3D \revisedPE{NLTE} Ba and Eu abundances, whether or not the CEMP subgroup classification is still meaningful in 3D. Because the abundance corrections themselves depend on metallicity, it is not straightforward to assume that a separation based on a gap in the 1D $\rm A(C)$ will still be present in 3D.

Integrating CEMP-r/s and CEMP-no distributions, we find that there are overall fewer CEMP stars in 3D. Below $\rm [Fe/H]=-4$ our sample contains $95\%$ CEMP stars in 1D, while in 3D this number reduces to $81\%$. Between $\rm [Fe/H]=-3.5$ and $\rm [Fe/H]=-3.0$ we consistently find a cumulative fraction that is $\sim 20\%$ lower in 3D than in 1D.
Above $\rm [Fe/H]=-3$ the 3D corrections in general are more modest, which is why we find that cumulative fractions agree to below $10\%$.

If one considers CEMP fractions in metallicity bins rather than as cumulative sum of stars below a given metallicity, it is straightforward to estimate the misclassification of CEMP stars around a specific metallicity. Table \ref{tab:cemp_frac} lists the fraction of CEMP stars using the same binning as in Fig. \ref{fig:cfe}. In total, we find that $\rm \sim 30\%$ of the CEMP stars in the range $\rm -4.5 < [Fe/H] <-3.5$ are not carbon-enhanced enough to qualify as such. In the ranges $\rm -3.5 < [Fe/H] <-2.5$ and $\rm -5.5 < [Fe/H] <-4.5,$ the misclassification is still $15\%$ and $12\%$, respectively.

In summary, we conclude that taking 3D CEMP effects on the C abundances into account reduces the observed CEMP fraction by $\sim 20\%$ in the metallicity regime below $\rm [Fe/H]=-3.0$. At the same time, the relative contribution of CEMP-no stars with respect to CEMP-r/s stars increases. Below $\rm [Fe/H]\approx-4$ we find that, in 3D, more than $80\%$ of CEMP stars are CEMP-no stars.

\subsection{Comparison with the literature}

The total CEMP fractions derived using 3D models are in good agreement with 1D-model based CEMP fractions reported in \citet{Placco2014}, but are somewhat lower in our distributions. \citet{Placco2014} find a fraction of $81\%$ CEMP stars at metallicities below $\rm [Fe/H]=-4$ and $100\%$ below $-5$, which is in good agreement with our findings. At $\rm [Fe/H]=-3.5$ and $\rm [Fe/H]=-3.0$, \cite{Placco2014} report a fraction of $60\%$ and $43\%$, respectively. These values are based on MS, sub-giants, and giants (their Fig. 4). In our 3D-corrected [C/Fe] distributions based on MS and sub-giants, the fractions are $\approx 45\%$ at $\rm [Fe/H]=-3.5$ and $\approx 33\%$ at $\rm [Fe/H]=-3.0$, which results in a 3D CEMP effect similar to what we find in this study. 

We note that \citeauthor{Placco2014} also explored the impact of 3D on C abundances in giants. Using the 3D line formation calculations for CH by \citet{Collet2007}, they estimate that negative 3D corrections, in giants with $\rm log(g)<3$, of $-0.3$ for $\rm -2.5 < [Fe/H] < -2.0$, $\rm -0.5$ for $\rm-3.0 <[Fe/H] < -2.5$, and $-0.7$ for $\rm[Fe/H]=-3.0,$ may arise. These estimates are similar to our results; however, our analysis confirms that departures from 1D HE are also prominent in the CH G-band spectra of MS stars. We note that the effect of C-enhancement in the 3D RHD atmosphere itself was not taken into account by \cite{Collet2007} and, hence, was not included in the 3D estimates by \cite{Placco2014}. As we show in this work, 3D effects are relevant for stars in all evolutionary phases. They thus can have a more significant impact on the overall distribution of CEMP stars across all metallicities than \citeauthor{Placco2014} anticipated.

\begin{figure}
    \centering
    \includegraphics[width=\columnwidth]{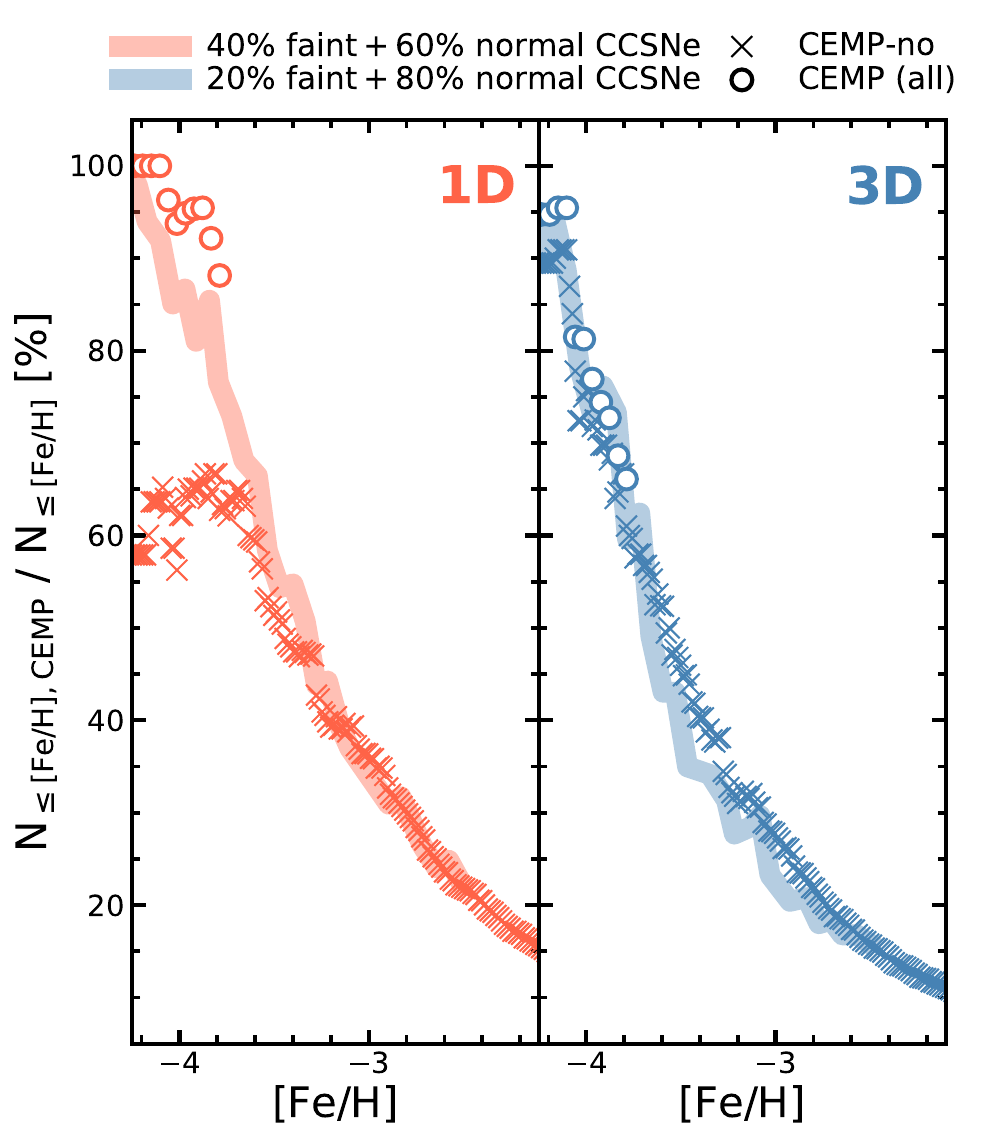}
    \caption{Cumulative CEMP-no fraction predictions from \cite{Hartwig2018} compared to our data. Below $\rm [Fe/H] < -3.75$, all CEMP stars in our sample are shown (circles). 3D CEMP data are shown in blue (right panel) and 1D data in red (left panel). The red line in the left panel corresponds to the fiducial model of \cite{Hartwig2018}, which comprises $40\%$ faint \citep[from][]{Ishigaki2014} and $60 \%$ normal  CCSNe \citep[from][]{Nomoto2013}. The blue curve in the right panel corresponds to their $20\%$ faint, $80 \%$ normal CCSN model.}
    \label{fig:hartwig}
\end{figure}
Since we find lower CEMP fractions at all metallicities, a question arises: what are the implications of lower CEMP occurrence rates in the early Galaxy? \citet{Hartwig2018} investigated the formation of the first generation of stars using a semi-analytical model\footnote{An early version of the A-SLOTH model; see \citet{Hartwig2022,Hartwig2024}.} based on dark matter halo merger trees that yield Milky Way-like halos. They probed the influence of various input parameters to their simulations on the observed CEMP occurrence rate as a function of [Fe/H] and compared their predictions to the data of \cite{Placco2014}. Faint SNe were considered as the primary channel for the formation of CEMP-no stars \citep[e.g.][references therein]{Klessen2023}. These objects, although spanning the same mass range as normal CCSNe, produce little $^{56}$Ni because of low explosion energy and/or effects of geometry in the explosion \citep{Tominaga2007}. Faint Pop III SNe generally produce a higher $\rm [C/Fe]$ ratio, because Fe-peak elements fall back onto the remnant and are not ejected. In the fiducial model, which is tuned to reproduce the 1D CEMP data from \cite{Placco2014}, \citet{Hartwig2018} include $40\%$ of faint \citep{Ishigaki2014}, and $60\%$ normal CCSNe \citep{Nomoto2013}. In addition, \cite{Hartwig2018} provide predicted CEMP fractions for a model that includes significantly fewer faint SNe (only $20\%$ instead of $40\%$). A comparison of specifically those two models with the data presented in this work can be found in Fig. \ref{fig:hartwig}. We note that \citeauthor{Hartwig2018} do not account for r-process and s-process enrichment in their model and hence only provide predictions for the fraction of CEMP-no stars. As mentioned in Sect. \ref{subsec:cemprs}, the CEMP-r/s or CEMP-no classification without s- and r-process abundance measurements is a relatively uncertain endeavour. At low metallicities, the rigid A(C) cut especially can result in a bias by wrongfully assigning high [C/Fe] group \uproman{3} stars as CEMP-r/s (see Fig. \ref{fig:yoon-beers}), even though group \uproman{3} predominantly consists of CEMP-no stars \citep[e.g.][]{Yoon2016}. Motivated by Fig. \ref{fig:yoon-beers}, we chose to include all CEMP stars below $\rm[Fe/H] < -3.75$ -- corresponding to the low-metallicity end of the blue group \uproman{1} ellipse -- and only CEMP-no stars at higher metallicities in the comparison with \citet{Hartwig2018} shown in Fig. \ref{fig:hartwig}. This effectively assumes that group \uproman{3} is composed exclusively of CEMP-no stars. We note that this assumption does not affect the overall conclusions, because we are mainly focusing on the shift between 3D CEMP and 1D and the metallicity region above $-3.5$. For a more detailed investigation, s- and r-process abundance measurements are essential.

Remarkably, their model with a $20\%$ faint SN fraction (solid blue line) is in excellent agreement with our 3D-corrected CEMP fractions (blue crosses + circles), while their fiducial model is in good agreement with our 1D data (in red). We additionally note that the systematic offset between 3D CEMP and 1D data at metallicities above $-3$ resembles the systematic offset between the fiducial and $20\%$ faint SN \citeauthor{Hartwig2018} models very well. At intermediate to low metallicities the CEMP-r/s -- CEMP-no classification becomes less precise due to the mixing of different CEMP groups. Nevertheless, because of the higher CEMP-no contribution in 3D, the agreement between model and observations is excellent also below $\rm [Fe/H]=-3.75$, regardless of whether the assumption of nearly $100\%$ CEMP-no stars in group \uproman{3} holds. In 1D, this is not the case; only after including all CEMP group \uproman{3} stars is the agreement restored, indicating that the CEMP-no classification is the culprit here. However, independent of the agreement at low metallicity in 1D, Fig. \ref{fig:hartwig} shows clearly that 3D-correction induced shifts in the CEMP fraction can be met with a decrease in the contribution of faint SNe in GCE. \cite{Hartwig2018} furthermore note that the $20\%$ faint SN model produces the approximately same metallicity distribution function for the halo as their fiducial model, and hence remains in good agreement with observations. An overestimation of the CEMP fraction thus leads to an overestimation of the contribution from faint SNe in the evolution of the early Galaxy. This is very interesting, as faint SNe will also impact the chemical enrichment of other chemical elements and hence large robust 3D \revisedPE{NLTE} abundance samples from next-generation facilities, like 4MOST \citep{Bensby2019}, will be in position to constrain the contribution of faint SNe to the GCE.
\section{Conclusions} \label{sec:conclusions}

In this work we show that physically realistic 3D RHD models of stellar atmospheres are crucial for the determination of accurate abundances of carbon in stars. We also show that they have a direct impact on our understanding of nucleosynthesis and the GCE.

We explore the effects of 3D convection on the critical diagnostic G band in spectra of MS and sub-giant stars. The G band arises due to absorption by the CH molecule and provides the strongest observational constraint on C abundances and [C/Fe] ratios in metal-poor stars. To compute 3D RHD simulations of stellar sub-surface convection, we used our new code, \texttt{M3DIS}, which was presented in \citet{Eitner2024}. The uppermost parts of these models were used to calculate detailed 3D spectrum synthesis around the G band and to derive a grid of 3D abundance corrections to [C/Fe] for the entire range of metallicities relevant for studies of GCE, $-6 \leq$ [Fe/H] $\leq -2$. The 3D models were computed for the first time using C-enhancements in line with low-metallicity observations. We compared the 3D synthetic spectra in the G band with the equivalent simulations computed using classical 1D HE models and quantified the amplitude of 3D effects on stellar [C/Fe] abundances.

We find that 3D effects on the G band are substantial and lead to lower carbon abundance estimates. This effect arises due to cooler outer structures of 3D RHD models compared to 1D HE structures. The formation of molecules is much more efficient in cooler 3D models, which implies that the spectral lines of CH are stronger. As a consequence, a lower abundance of C is required to fit a given observed spectrum, and the 3D corrections are negative. Interestingly, increasing C abundances in 3D models at the same metallicity (set by iron) yields the opposite effect: the G band weakens. Physically, this is because increasing C (to levels observed in CEMP stars) implies a higher radiative opacity in the models and leads to slightly hotter 3D models compared to their scaled-solar counterparts. The 3D CEMP A(C) corrections for MS stars \revisedPE{can} exceed $\rm A(C)_{3D\ CEMP} - A(C)_{1D} =-0.6\ dex$ at $\rm [Fe/H]=-5$. For sub-giants, the 3D CEMP effects are of the order of $\rm\sim -0.7$ dex, in good agreement with previous estimates in the literature \citep{Collet2018, Gallagher2016,Gallagher2017}. The differences between 3D and 1D A(C) decrease with increasing [Fe/H].

We applied 3D CEMP A(C) abundance corrections to a sample of metal-poor stars from the SAGA database to investigate the impact on the Galactic CEMP population. We find a significant fraction of CEMP stars across all metallicities that are no longer classified as C-enhanced. In the range $\rm -4.5 < [Fe/H] < -3.5$ we find $\sim 30\%$ fewer CEMP stars after applying the 3D abundance corrections, and in the range $\rm -3.5 < [Fe/H] < -2.5$ the number is reduced by $15\%$. We furthermore find that the cumulative CEMP fraction below $\rm[Fe/H]=-3$ is $\rm \sim 20\%$ lower in 3D than it appears in 1D. At lower metallicities, the distribution remains mostly unchanged because the same $\rm A(C)$ corresponds to a higher $\rm [C/Fe]$. As a consequence, the $\rm A(C)$ plateau remains present in 3D, though it appears to be tilted by $10^{\circ}$. When considering the CEMP-no and CEMP-r/s subgroups based on $\rm A(C)$ \citep{Yoon2016}, we find a reduction in the CEMP-r/s sample size in 3D, as a significant number of stars are shifted either below the CEMP limit of $\rm[C/Fe]=0.7$ or below the CEMP-r/s limit of $\rm A(C)=7.1$. Consequently, the use of 3D-corrected C abundances leads to a higher fraction of CEMP-no stars, as some stars previously classified as CEMP-r/s are re-classified. Despite their generally lower metallicities, CEMP-no stars retain a high $\rm A(C)$, ensuring that most remain classified as carbon-enhanced.

The fraction of CEMP stars as a function of metallicity is a tracer of the contribution of faint SNe to the GCE \citep{Hartwig2018, Klessen2023}. Following the models of \cite{Hartwig2018}, we find that a lower fraction (only 20\%) of faint SNe in the Pop III galaxy formation models can describe the 3D CEMP distributions. This is an interesting result that remains to be confirmed with detailed GCE models and other chemical elements, which are sensitive to faint SN contributions.

\section*{Data availability}
A table with references, stellar parameters, and [C/Fe] abundances is only available in electronic form at the CDS via anonymous ftp to \url{cdsarc.u-strasbg.fr} (130.79.128.5) or via \url{http://cdsweb.u-strasbg.fr/cgi-bin/qcat?J/A+A/}

\begin{acknowledgements}
MB is supported through the Lise Meitner grant from the Max Planck Society. PE is supported through the IMPRS fellowship of the Max Planck Society.

AP work is supported by the Research Council of Norway through its Centres of Excellence scheme, project number 262622, as well as through the Synergy Grant number 810218 (ERC-2018-SyG) of the European Research Council.

RSK acknowledges financial support from the ERC via Synergy Grant ``ECOGAL'' (project ID 855130),  from the German Excellence Strategy via the Heidelberg Cluster ``STRUCTURES'' (EXC 2181 - 390900948), and from the German Ministry for Economic Affairs and Climate Action in project ``MAINN'' (funding ID 50OO2206).  RSK also thanks the 2024/25 Class of Radcliffe Fellows for highly interesting and stimulating discussions. 

Computations were performed on the HPC system Raven at the Max Planck Computing and Data Facility. 
\end{acknowledgements}

\bibliographystyle{aa}
\bibliography{aanda.bib} 

\appendix
\section{Model parameter selection}\label{appendix:model_parameter_selection}
\begin{figure}
    \centering
    \includegraphics[width=\columnwidth]{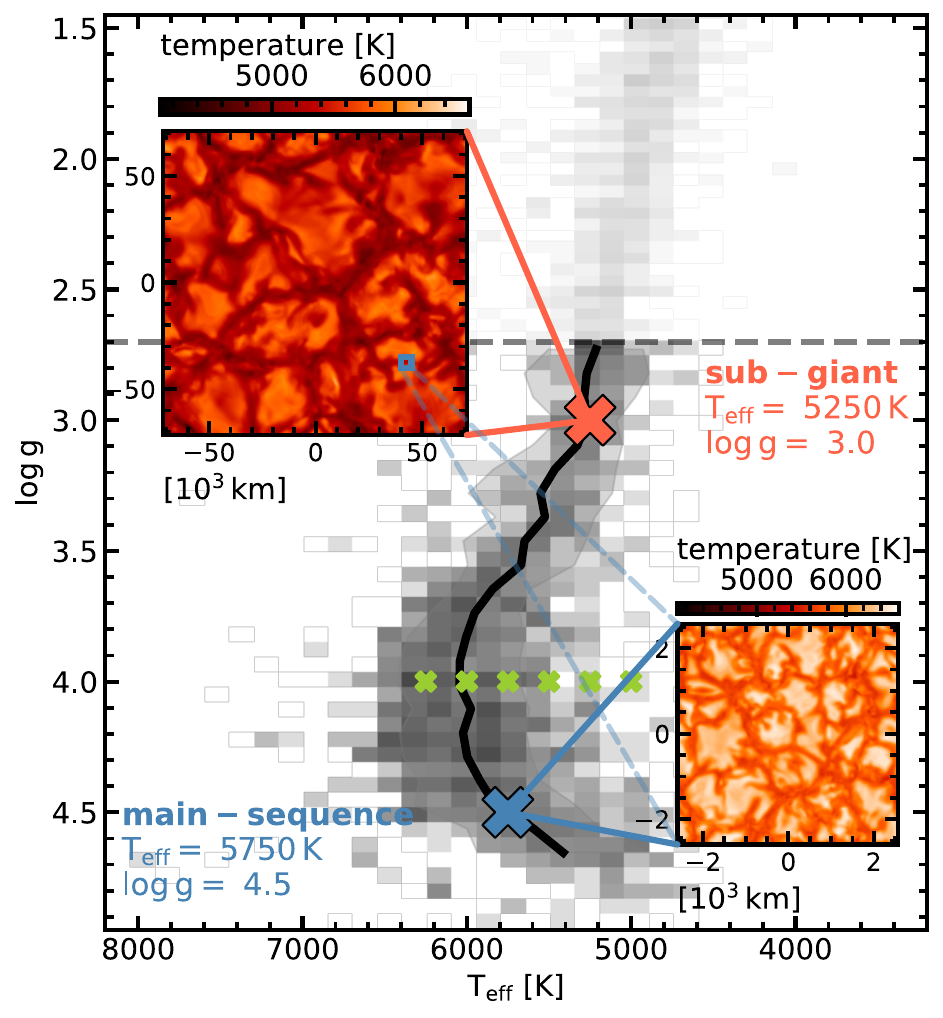}
    \caption{2D histogram of the dataset used in this paper in the $\rm T_{eff}$ vs $\rm \log g$ plane. The $\rm \log g$-limit of $2.7$ applied in this work is shown as a dashed black line. The solid black line represents the mean in bins of constant $\rm \log g$. Red and blue crosses mark the location of the 3D \medis\ models. The insets show the temperature distribution at the optical surface of the respective model with $\rm [Fe/H]=-5$. To highlight the different physical scales involved, the blue square in the inset shows the size of the MS model. To estimate the dependence of 3D $\rm A(C)$ corrections on effective temperature, additional models are created at the locations highlighted with green crosses.}
    \label{fig:hist_model_selection}
\end{figure}
\begin{figure}
    \centering
    \includegraphics[width=\columnwidth]{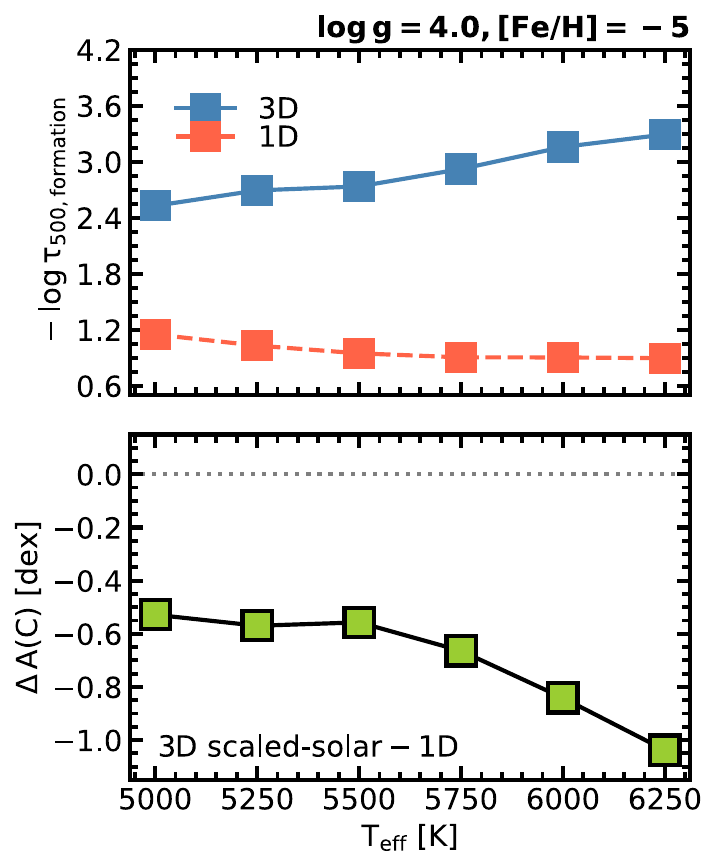}
    \caption{Top: Formation height of a representative CH line in the atmospheres of 3D \medis\ models at $\rm [Fe/H]=-5,\ \log g=4.0$. Bottom: 3D scaled-solar -- 1D $\rm A(C)$ abundance corrections as a function of $\rm T_{eff}$, derived from spectra using the same atmosphere models as in the top panel and applying $\rm [C/Fe]=3.0$ as the reference abundance, as in Fig. \ref{fig:corrections}}
    \label{fig:abund_corr_teff}
\end{figure}

To justify the selection of model parameters, we show the $\rm T_{eff}$-$\rm \log g$ distribution of the stars in our dataset in Fig. \ref{fig:hist_model_selection}. Given the numerical complexity of the analysis in the paper, we restricted the number of parameters and constructed 3D \medis\ model atmospheres for these two $\rm T_{eff}$-$\rm \log g$ combinations only (see Table \ref{tab:models}), which are chosen to best represent the stellar sample. To quantify the influence of this selection on the $\rm 3D-1D$ abundance corrections, we additionally computed a range of 3D \medis\ models in between, with a single representative chemical composition of $\rm [Fe/H]=-5, [C/Fe]=0$ and $\rm \log g=4.0$ (green crosses). We created these models for a range of different effective temperatures between $\rm 5000\,K$ and $\rm 6250\,K$ to estimate the effect of $\rm T_{eff}$ on abundance corrections, while keeping $\rm \log g$ fixed. The results are shown in the bottom panel of Fig. \ref{fig:abund_corr_teff}.

When comparing this panel with Fig. \ref{fig:corrections}, we see that the $\rm 3D\ \text{scaled-solar} - 1D $ $\rm A(C)$ abundance corrections overall are very similar to the corrections computed for MS and sub-giant models, indicating that there is no significant effect on the overall distribution of CEMP stars that influences the conclusions drawn in this paper. At the same $\rm T_{eff}=5750\,K$, the difference between the MS ($\rm \log g = 4.5$) and the additional model ($\rm \log g = 4.0$) is negligible. Corrections for the sub-giant are slightly larger than the results at $\rm T_{eff}=5250\,K$, which is expected and can be attributed to its significantly lower surface gravity. Stars near $\rm \log g=4.0$ align more closely with the MS model than with the sub-giant model regarding stellar parameters, so this discrepancy is not relevant for the final distribution.

Investigating the corrections as a function of effective temperature, it becomes apparent that $\rm \Delta A(C)$ increases (in absolute values) as $\rm T_{eff}$ increases. Considering that the strength of molecular spectral lines weakens with increasing temperature, this may seem counter-intuitive at first. To understand this behaviour qualitatively, we additionally include the formation height of a representative CH line in the G band for different $\rm T_{eff}$ in the upper panel of Fig. \ref{fig:abund_corr_teff}, in 1D (dashed red) as well as in 3D (solid blue). Investigating the 1D trend, we see that -- as expected -- the formation height of CH lines decreases, i.e. shifts towards the stellar interior, as the atmosphere gets hotter. However, in 3D the opposite is the case; here the negative temperature gradient in the upper layers of the atmosphere -- which is not present in 1D models in radiative equilibrium -- still provides conditions favourable for molecule formation in cooler layers further out, even though the overall temperature at the optical surface increases. The formation height hence shifts in the opposite direction, which causes the structural and spectral differences between 3D and 1D to increase. Because the hottest model we included in the main analysis has $\rm T_{eff}=5750\,K$, we find that the $\rm A(C)$ correction for stars in our sample around $\rm \log g \sim 4.0,\ [Fe/H]\sim-5$ at the location of the mean ($\rm T_{eff}\sim 6000\,K$) may be underestimated by $\sim\rm 0.2\, dex$. We thus expect that 3D effects become even more significant when the sampling of the parameter space is increased, which further supports the conclusions drawn in this study and underlines the need for 3D model atmospheres for the analysis of metal-poor stars. It remains to be investigated how this temperature dependence behaves as a function of metallicity and with C-enhancement included in the computation of the atmosphere itself.

\begin{table}
\caption[]{CEMP fractions in bins of $\rm \Delta[Fe/H]= \pm 0.5\ dex$.}  
\label{tab:cemp_frac}  
\centering
\begin{tabular}{crr}  
\hline\hline                 
$\rm [Fe/H]$ & 1D & 3D \\  
$\rm dex$ & $\%$ & $\%$ \\ 
\hline                        
-6 & 100 & 100 \\
-5 & 100 & 88 \\
-4 & 83 & 53 \\
-3 & 50 & 35 \\
-2 & 26 & 19 \\
-1 & 19 & 12 \\
\hline                                   
\end{tabular}
\end{table}
\section{Updated radiative transfer solver}\label{appendix:supp_material}

\begin{figure}
    \centering
    \includegraphics[width=\columnwidth]{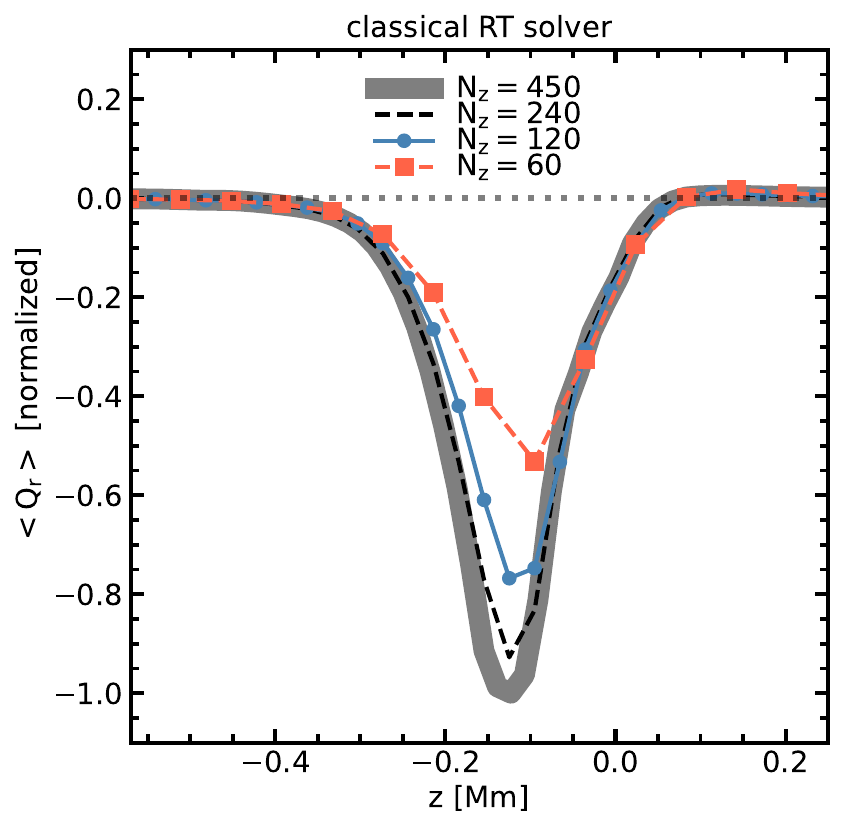}
    \includegraphics[width=\columnwidth]{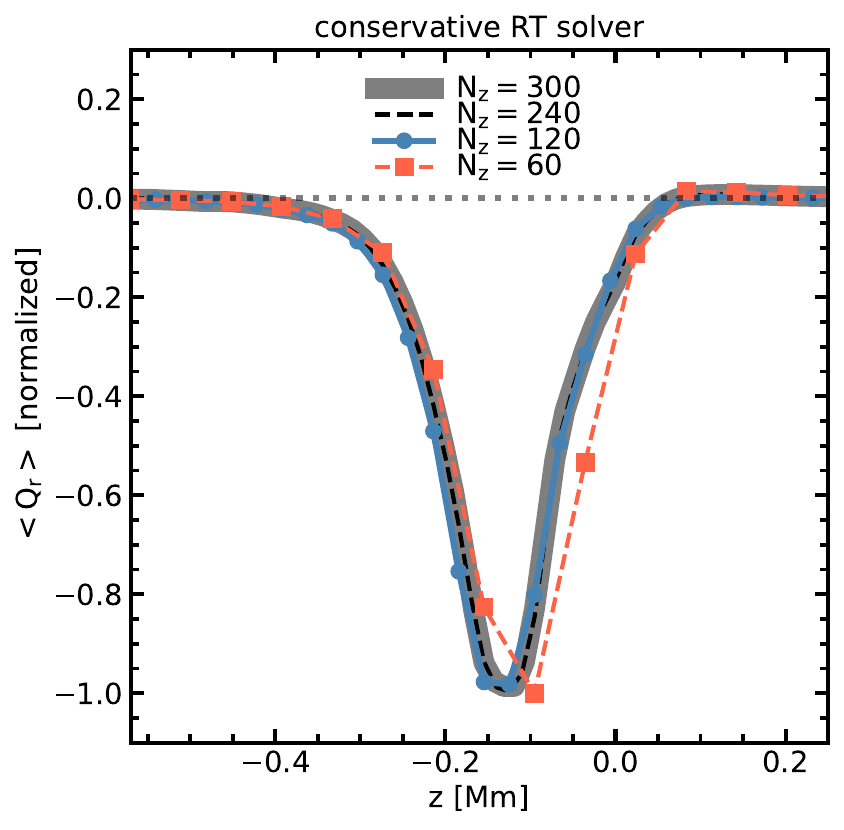}
    \caption{Comparison of the horizontally averaged radiative heating ($\rm Q_r$), computed at different resolutions for the same model atmosphere using the old, classical RT solver (top) and the new, conservative RT solver on a staggered grid (bottom). The new solver already reproduces the cooling peak very well at intermediate resolution, while the classical solver shows the typical under-sampling effect, which leads to an artificial heating of the model.}
    \label{fig:newheatingsolver}
\end{figure}
We have found that the cost of computing accurate vertical temperature profiles at the surface can be significantly reduced, relative to \cite{Eitner2024}, by modifying the centring of RT solutions in the short-characteristics solver. Instead of computing the radiation intensities $\rm I_{\Omega,\nu}$ and source functions ($\rm S_{\nu}$) at cell centres, and computing the net radiative heating from
\begin{equation}
    Q_{\rm r} = \iint {\rm d}\Omega\, {\rm d}\nu \, \rho\kappa_{\nu}\, (I_{\rm \Omega,\nu}-S_{\rm \nu})\,,
\end{equation}
one can instead compute the radiative intensities $\rm I_{\Omega\nu}$ at cell interfaces, and from these compute the radiative fluxes $F_{i,\nu}$ in each direction $i$, which allows the heating to be obtained from
\begin{equation} 
    Q_{\rm r} = -\int {\rm d}\nu \ \nabla\cdot\mathbf{F_{\rm \nu}}\,,
\end{equation}
where the flux divergence is computed numerically as the differences between incoming and outgoing fluxes for each cell. This method is by construction energy-conserving, in the sense that the cooling corresponds exactly to the energy increase in the radiation field leaving the cells, relative to the energy that was entering. Thus, while the previous method `samples' the cooling point-wise, the improved method computes the actual energy loss across cells.

The improvement is illustrated in Fig.\ \ref{fig:newheatingsolver}, which plots horizontally average cooling profiles for different vertical resolutions.
\end{document}